%% file: Gatewayportal.tex
\documentclass[11pt,a4paper]{article}
\usepackage[T1]{fontenc}
\usepackage[utf8]{inputenc}
\usepackage{lmodern,microtype}
\usepackage[margin=26mm]{geometry}
\usepackage{amsmath,amssymb,bm,mathtools}
\usepackage{booktabs,graphicx,placeins}
\usepackage{pgfplots}
\pgfplotsset{compat=1.18}
\usepackage[numbers,sort&compress]{natbib}
\usepackage[colorlinks=true,linkcolor=blue!50!black,citecolor=blue!50!black,urlcolor=blue!50!black]{hyperref}
\allowdisplaybreaks
\newcommand{\GeV}{\mathrm{GeV}}
\newcommand{\eV}{\mathrm{eV}}
\newcommand{\mpl}{\overline M_{\mathrm{Pl}}}
\newcommand{\dd}{\mathrm d}
\newcommand{\tr}{\operatorname{tr}}
\title{\large\bfseries A Gauge Pathway to an Ultralight Axion through a Hidden Portal}
\author{Claudio Corian\`o\\[5pt]
\small Dipartimento di Matematica e Fisica ``Ennio De Giorgi'', Universit\`a del Salento,\\
\small INFN, Sezione di Lecce, and CNR NANOTEC, Lecce, Italy}
\date{}
\begin{document}
\maketitle
\begin{abstract}
We study a gauge realization of an ultralight axion with a hidden $SU(2)$ interaction, an Abelian vector portal, a charged Higgs and a St\"uckelberg field. The vector absorbs one scalar combination and leaves a physical phase. Its normalization is fixed by the Higgs and St\"uckelberg scales, while hidden confinement generates its mass. Keeping this phase ultralight requires an additional symmetry: the displayed gauge charges alone allow large phase-dependent scalar interactions. We quantify the suppression required of these interactions. Pure-$SU(2)$ lattice results relate the susceptibility to physical confinement scales; one- and two-loop running illustrate the dependence on the chosen high-energy coupling. We derive the coherent radiation-era density and exhibit an initial angle giving the reference dark-matter abundance for $F_\chi\simeq10^{17}\,\GeV$ and zero-temperature mass $m_{\chi,0}=3\times10^{-20}\,\eV$. We follow the confined population through number-changing and number-conserving limits and bound its energy contribution. An illustrative inflaton background satisfies the stated vacuum and isocurvature tests, conditional on control of a large portal correction to its potential. Its cold hidden history requires a hidden-to-visible reheating fraction near $5\times10^{-12}$, whose microscopic origin remains an input. A second, gauge-anomaly-free model gives the same physical phase scale using two charged Higgs singlets. These constructions specify conditional cosmological examples and the additional protection and reheating mechanisms needed to complete them.
\end{abstract}
\clearpage

\section{Introduction}
\label{sec:intro}
Cold dark matter describes large cosmological scales well, while its behaviour within galaxies remains an important test of its microscopic properties. The central densities of halos, the number of small halos and the motions of stars in dwarf galaxies have motivated changes to the collisionless particle picture. Gas dynamics, star formation and the relation between observed galaxies and their host halos also affect these comparisons. An ultralight bosonic field changes gravitational clustering when its de Broglie wavelength becomes comparable to a halo core. Field gradients and wave interference then affect the distribution of dark matter \cite{ULA}. The same wave effects reduce the growth of short-wavelength density perturbations. Observations, including the Lyman-$\alpha$ forest, therefore constrain both the axion mass and the fraction of dark matter that it supplies \cite{Lyman}.

A portal connects Standard Model particles to a hidden sector through interactions that can produce hidden particles and transfer energy between the two sectors. The hidden sector can contain several forms of dark matter: a coherently oscillating ultralight field, massive fermions or bound states of a hidden gauge force. Their abundances and scattering rates depend on the particle spectrum and its cosmological history. The portal can also mediate scattering among dark particles, providing a framework for considering both nearly collisionless and interacting components. This possibility motivates the present study. We use a massive Abelian vector coupled to visible baryon minus lepton number and to a hidden charge, and investigate the gauge origin and abundance of an ultralight component together with the constraints on accompanying hidden populations.

String theory provides one origin for light axions. Antisymmetric tensor fields can give four-dimensional axions when their components are integrated over closed cycles in the extra dimensions. The geometry determines the axion kinetic terms and the size of effects such as instantons that generate their potentials \cite{StringAxions,Axiverse}. In many models, the axion shares a scalar multiplet with fields called moduli, which describe sizes or shapes of the compact dimensions. Finding stable values for the moduli is then part of determining which axions remain light. Here we study a four-dimensional gauge theory in which the scalar phases, their interactions and the scales entering the axion mass are specified directly.

The conventional QCD axion illustrates an important restriction on the relation between mass and interactions. In the Peccei--Quinn (PQ) mechanism, the axion is the angular mode associated with spontaneous breaking of an approximate global symmetry \cite{PQMechanism}. When QCD dominates its potential, the axion decay constant $f_a$ fixes its zero-temperature mass through
\begin{equation}
 m_a^2f_a^2=\mathcal X_{\rm QCD}(0),\qquad
 m_a\simeq5.7\,\mu\mathrm{eV}
 \left(\frac{10^{12}\,\GeV}{f_a}\right).
 \label{eq:qcd-axion-mass-relation}
\end{equation}
Here $\mathcal X_{\rm QCD}(0)$ is the curvature of the QCD vacuum energy with respect to its theta angle. The same scale suppresses the interactions. For example, the axion--photon coupling is
\begin{equation}
 g_{a\gamma\gamma}=\frac{\alpha_{\rm em}}{2\pi f_a}C_\gamma,
 \label{eq:qcd-axion-photon-relation}
\end{equation}
where $\alpha_{\rm em}$ is the electromagnetic fine-structure constant and $C_\gamma$ is a dimensionless coefficient determined by the charge assignments and QCD mixing \cite{QCDPrecise}. Couplings to ordinary fermions also involve the inverse scale $f_a$, with model-dependent coefficients. Once these coefficients are fixed, a larger mass implies stronger interactions.

This correlation explains how limits on interactions become bounds on the QCD-axion mass. Axions produced in stellar interiors can carry energy away and change the observed stellar lifetimes, constraining their couplings \cite{AxionStellarCooling}. Searches for solar axions also bound their conversion into photons \cite{SolarAxionSearch}. Through the common scale $f_a$, such coupling bounds restrict the mass in a specified QCD-axion model. Cosmological production supplies additional restrictions: for a specified initial angle and thermal history, the relic abundance must not exceed the observed dark-matter density \cite{QCDPrecise}. The resulting mass intervals therefore depend on the interaction coefficients and the assumed cosmological history.

A complementary route constructs axion-like particles through an extended gauge and Higgs sector. A St\"uckelberg pseudoscalar, which shifts under an additional Abelian gauge transformation, mixes with phases of charged Higgs fields. Some combinations supply the longitudinal polarizations of massive gauge bosons, while an independent gauge-invariant combination can remain as a physical pseudoscalar. This surviving mixed state is called the \emph{axi-Higgs}. Its composition and couplings follow from the gauge charges, the Higgs vacuum values and the St\"uckelberg scale; its mass depends on the potential for the surviving angle.

In anomalous Abelian gauge extensions, interactions involving the shifting St\"uckelberg field cancel the anomalous gauge variation of the fermion quantum action. The resulting axion--gauge couplings depend on the St\"uckelberg scale and on the amount of that field contained in the physical axi-Higgs. The strength of its periodic potential is an additional input, allowing a mass--coupling relation different from that of the QCD axion \cite{GaugedAxionMassCouplings}. Consequently, mass bounds obtained by assuming Eqs.~\eqref{eq:qcd-axion-mass-relation} and \eqref{eq:qcd-axion-photon-relation} do not apply unchanged. Stellar, experimental and cosmological constraints must instead be evaluated with the physical couplings and production history. A surviving coupling to the QCD topological density would still generate a QCD contribution to the potential. In the models studied here, the intended ultralight mass is generated by the hidden non-Abelian interaction.

Models based on $E_6\times U(1)$ provide explicit realizations of this mechanism in which a periodic potential suppressed by high-scale instantons can make the axi-Higgs ultralight \cite{GUTAxion,StAxions}. Vector-portal models realize the mixing with a charged hidden Higgs and a St\"uckelberg field, together with interactions that cancel the gauge anomalies \cite{Portal}. We build on this phase-mixing mechanism while using a simpler hidden gauge group and studying the ultralight population through its production history.

The St\"uckelberg field has a tensor description. In four dimensions, a massless antisymmetric two-form field has one propagating degree of freedom and can be described by a pseudoscalar. A tensor--vector coupling gives the St\"uckelberg mass term after duality, as occurs in effective actions obtained from string compactifications \cite{OrientifoldAction,StAxions}. A separate field-theory realization retains the phase of a charged scalar after integrating out its heavy radial excitation \cite{ScalarDecoupling}. Section~\ref{sec:stueckelberg-origin} distinguishes these constructions and states the scalar assumptions used in our inflationary example. In either description, mixing with the hidden Higgs phase leaves a physical axi-Higgs whose potential is generated by hidden gauge dynamics.

Compared with the $E_6\times U(1)$ realizations, our construction uses scalar singlets and fermions in singlet or doublet representations of a hidden $SU(2)$. The hidden Higgs leaves this non-Abelian group unbroken, allowing it to confine at a scale far below the Higgs expectation value. The large scalar scale fixes the normalization of the physical phase, while the low confinement scale supplies its small angular curvature. We thus examine a different origin for the small potential from the high-scale instanton construction. Building on the portal framework of Ref.~\cite{Portal}, we combine this separation of scales with a calculation of coherent axion production and an assessment of the accompanying confined population. We also give a separate two-Higgs realization whose displayed fermion spectrum cancels all gauge anomalies.

Keeping the phase ultralight requires suppression of additional phase-dependent interactions. Gauge invariance allows scalar operators of dimension two that would otherwise give a large angular mass. We impose a perturbative phase symmetry, quantify the required protection, and state the remaining high-energy assumptions. Constructions in which gauge charges also restrict the leading symmetry-breaking operator address this protection by a different choice of charges and field content \cite{GaugedPQ}.

The hidden gauge coupling and initial displacement can be selected to obtain the reference dark-matter density. Pure-gauge lattice results relate the angular curvature to physical confinement scales, while the conversion to a high-energy coupling carries running and threshold uncertainties. Hidden glueballs, the bound states produced at confinement, have their own number-changing reactions and relic abundance \cite{Glueballs}. We follow their evolution and bound their contribution alongside the coherent axion. A conditional inflaton example tests the survival of the scalar vacuum, isocurvature fluctuations and the reheating conditions needed for the chosen history.

An interacting component would open further possibilities for structure formation. Collisions can carry heat out of a halo core and drive gravothermal contraction, potentially leading to black-hole seeds \cite{CollapseSeeds,EarlySeeds,BaryonicSeeds}. Connections to little red dots also require accretion and surrounding gas \cite{LRDSeeds,Accretion}. The examples calculated here supply an almost collisionless coherent axion and constrain accompanying relics. A populated component with efficient halo scattering requires suitable lighter particles and a mediator, together with its own production and transport calculation. We state the conditions for that extension in the outlook.

\section{Gauge action and the physical phase}
\label{sec:model}
We extend the Standard Model by a hidden sector whose non-Abelian gauge dynamics can generate an axion potential. A shared Abelian vector connects this sector to visible matter. The hidden scalar sector contains a charged Higgs and a St\"uckelberg pseudoscalar. When the Higgs develops a vacuum expectation value, its phase mixes with the St\"uckelberg field. One combination supplies the longitudinal polarization of the massive Abelian vector, leaving the physical axi-Higgs. We introduce the gauge symmetry and field content below, then derive this surviving field and its normalization from the scalar kinetic terms.

The gauge group and its two independent new couplings are
\begin{equation}
 G_{\rm SM}\times SU(2)_D\times U(1)_X,
 \qquad g_D\ \hbox{for }SU(2)_D,\qquad g_X\ \hbox{for }U(1)_X.
 \label{eq:group}
\end{equation}
The subscript $D$ denotes the hidden gauge group. The shared Abelian generator is
\begin{equation}
 X=(B-L)_{\rm vis}+D_{\rm hid}.
 \label{eq:shared-X-generator}
\end{equation}
The Abelian vector $X_\mu$ couples to visible $B-L$ and to the hidden charge $D_{\rm hid}$. The visible gauge anomalies cancel when three right-handed neutrinos are included. The visible Higgs has zero portal charge. All the hidden fields are singlets under the Standard Model gauge groups.

The hidden fermion $F$ has two components under $SU(2)_D$, each with Dirac spinor structure. Both its left- and right-handed parts transform as hidden doublets. The Dirac fermion $\Psi$ is an $SU(2)_D$ singlet, so it has no hidden-colour interaction, but it carries the Abelian portal charge. The Higgs $\phi$ is also an $SU(2)_D$ singlet. Table~\ref{tab:charges} lists the representations and charges.

\begin{table}[htbp]
\centering
\small
\setlength{\tabcolsep}{5pt}
\renewcommand{\arraystretch}{1.18}
\begin{tabular}{@{}lcccccc@{}}
\toprule
Field & Spin & $SU(3)_c$ & $SU(2)_L$ & $Y$ & $SU(2)_D$ & $q_X$\\
\midrule
$F_L$ & $\tfrac12$ & $\mathbf1$ & $\mathbf1$ & $0$ & $\mathbf2$ & $+1$\\
$F_R$ & $\tfrac12$ & $\mathbf1$ & $\mathbf1$ & $0$ & $\mathbf2$ & $-1$\\
$\Psi_L$ & $\tfrac12$ & $\mathbf1$ & $\mathbf1$ & $0$ & $\mathbf1$ & $+1$\\
$\Psi_R$ & $\tfrac12$ & $\mathbf1$ & $\mathbf1$ & $0$ & $\mathbf1$ & $-1$\\
$\phi$ & $0$ & $\mathbf1$ & $\mathbf1$ & $0$ & $\mathbf1$ & $+2$\\
\bottomrule
\end{tabular}
\caption{Hidden fields before symmetry breaking. The columns give the spin, Standard Model representations, hidden $SU(2)_D$ representation and portal charge $q_X=D_{\rm hid}$. The labels $L,R$ denote the left- and right-handed parts of each Dirac field.}
\label{tab:charges}
\end{table}

For the charge-two Higgs, define $e_X\equiv2g_X$. Before choosing the vacuum, the leading Higgs potential is
\begin{equation}
 V_{\rm rad}(\phi)=-\mu_\phi^2\phi^\dagger\phi
 +\lambda_\phi(\phi^\dagger\phi)^2.
 \label{eq:radial}
\end{equation}
We take $\lambda_\phi>0$. The vacuum value and scalar masses are obtained from the full potential.

Including possible kinetic mixing with hypercharge, the Lagrangian is
\begin{equation}
 \begin{aligned}
 \mathcal L={}&\mathcal L_{{\rm SM}+\nu_R}
 -\frac14W^a_{\mu\nu}W^{a\mu\nu}-\frac14F_{X\mu\nu}F_X^{\mu\nu}
 -\frac{\epsilon}{2}F_{Y\mu\nu}F_X^{\mu\nu}\\
 &+\frac12(\partial_\mu b-M_1X_\mu)^2+|D_\mu\phi|^2
 +\bar F i\gamma^\mu D_\mu F+\bar\Psi i\gamma^\mu D_\mu\Psi\\
 &-\left(y_F\phi\bar F_LF_R+y_\Psi\phi\bar\Psi_L\Psi_R+\mathrm{h.c.}\right)
 -V_{\rm rad}+\mathcal L_{\rm WZ}.
 \end{aligned}
 \label{eq:lagrangian}
\end{equation}
The field $b$ is the shifting pseudoscalar. The parameter $M_1$ sets the St\"uckelberg contribution to the vector mass. The parameter $\epsilon$ measures kinetic mixing with the hypercharge field strength $F_Y$. The term $\mathcal L_{\rm WZ}$ contains the anomaly-cancelling interactions given below. Here $W_\mu^a$ are the three hidden gauge fields, $W^a_{\mu\nu}=\partial_\mu W_\nu^a-\partial_\nu W_\mu^a+g_D\varepsilon^{abc}W_\mu^bW_\nu^c$, and $F_X=\dd X$. The visible interaction includes $g_XX_\mu J_{B-L}^\mu$. For $T^a=\sigma^a/2$ and $P_{L,R}=(1\mp\gamma_5)/2$, the derivatives are
\begin{align}
 D_\mu\phi&=(\partial_\mu-ie_XX_\mu)\phi,\notag\\
 D_\mu F&=\left[\partial_\mu-ig_DT^aW_\mu^a
 -ig_XX_\mu(P_L-P_R)\right]F,\notag\\
 D_\mu\Psi&=\left[\partial_\mu-ig_XX_\mu(P_L-P_R)\right]\Psi.
 \label{eq:derivatives}
\end{align}
Both Yukawa interactions have zero total gauge charge. We choose real positive Yukawa couplings and include constant fermion phases in the hidden theta angle. A conserved singlet fermion number protects $\Psi$ against decay. We also impose a global visible $B-L$ symmetry to forbid the otherwise gauge-allowed coupling $\phi\nu_R^TC\nu_R$. We assume that these symmetries remain valid in the theory at higher energies.

We also write the phase-dependent interaction in terms of the original fields. The combination
\begin{equation}
 \mathcal Z=\phi\,e^{-ie_Xb/M_1}.
 \label{eq:covariant-phase-product}
\end{equation}
is unchanged by the portal gauge transformation: $\phi$ acquires the factor $e^{ie_X\omega(x)}$ and $b$ shifts by $M_1\omega(x)$, giving
\begin{equation}
 \mathcal Z\longmapsto
 e^{ie_X\omega}\phi\,e^{-ie_X(b+M_1\omega)/M_1}
 =\mathcal Z.
 \label{eq:phase-product-invariance}
\end{equation}
The magnitude $\phi^\dagger\phi$ is also gauge invariant. A single periodic interaction of the form considered in the preprint of Ref.~\cite{Portal} gives
\begin{equation}
 \begin{aligned}
 V_{\rm hid}^{(1)}(\phi,b)={}&
 -\mu_\phi^2\phi^\dagger\phi
 +\lambda_\phi(\phi^\dagger\phi)^2\\
 &-\left[\kappa_{\rm ang}\phi e^{-ie_Xb/M_1}
 +\kappa_{\rm ang}^*\phi^\dagger e^{ie_Xb/M_1}\right].
 \end{aligned}
 \label{eq:covariant-single-operator-potential}
\end{equation}
The coefficient $\kappa_{\rm ang}$ has mass dimension three. This expression is a possible low-energy form of an anomaly-generated potential, with its magnitude and radial dependence to be supplied separately. We use the general gauge vacuum energy below for the confining model. An arbitrary perturbative coefficient of this form is excluded by the phase-protection assumption discussed in Section~\ref{sec:phase-protection}.

We now define the potential generated by the hidden gauge dynamics. The parameter $\theta_D$ is the dimensionless vacuum angle of the hidden $SU(2)_D$ gauge theory. It multiplies the hidden topological density,
\begin{equation}
 \mathcal L_{\theta_D}=\theta_D Q_D,
 \qquad
 Q_D=\frac{g_D^2}{32\pi^2}W^a_{\mu\nu}\widetilde W^{a\mu\nu},
 \qquad
 \widetilde W^{a\mu\nu}=\frac12\varepsilon^{\mu\nu\rho\sigma}
 W^a_{\rho\sigma}.
 \label{eq:top-density}
\end{equation}
For a finite-action gauge configuration, the spacetime integral of $Q_D$ is an integer topological charge. The path-integral weight therefore depends on $e^{i\theta_D n}$, and $\theta_D$ is periodic modulo $2\pi$. It is a fixed parameter of the Lagrangian. A chiral redefinition of a massive hidden fermion can move a phase between its mass and $\theta_D$. Their invariant combination determines the physical gauge angle. In the present model this angle is
\begin{equation}
 \vartheta=\theta_D+\arg\!\left(\phi e^{-ie_Xb/M_1}\right).
 \label{eq:physical-hidden-angle-early}
\end{equation}
After symmetry breaking, this angle becomes the combination of $\theta_D$ and the physical axi-Higgs field derived below. For a constant scalar value $\rho_\phi\equiv\phi^\dagger\phi$ and a fixed value of $\vartheta$, let $\mathcal E_D(\rho_\phi,\vartheta)$ denote the hidden vacuum energy density. We assign its angle-independent perturbative contributions to the radial effective potential. The remaining gauge contribution includes confinement. We approximate its angular dependence by the leading periodic harmonic,
\begin{equation}
 \mathcal E_D(\rho_\phi,\vartheta)
 =\mathcal E_{D,0}(\rho_\phi)
 +\mathcal A_D(\rho_\phi)\,[1-\cos\vartheta],
 \qquad
 \mathcal E_{D,0}(\rho_\phi)
 \equiv\mathcal E_D(\rho_\phi,0).
 \label{eq:hidden-vacuum-energy-form}
\end{equation}
The subtraction convention fixes how the angle-independent energy is divided between perturbative terms, the confinement contribution and the additive vacuum constant \cite{VacuumEnergySubtraction}. Since $\phi$ is an $SU(2)_D$ singlet, its magnitude enters the gauge theory through the doublet mass $m_F(\rho_\phi)=y_F\sqrt{\rho_\phi}$. For $m_F$ much larger than the confinement scale, one-loop matching gives the following leading dependence at fixed high-energy gauge coupling \cite{HeavyDecoupling,HeavyMassScaling}:
\begin{equation}
 \begin{aligned}
 \Lambda_D(\rho_\phi)
 &=\Lambda_1
 \left(\frac{y_F\sqrt{\rho_\phi}}{\Lambda_1}\right)^{1/11},\\
 \mathcal E_{D,0}(\rho_\phi)
 &=\mathcal E_{\rm ref}-c_E\Lambda_D^4(\rho_\phi)
 =\mathcal E_{\rm ref}-c_E\Lambda_1^4
 \left(\frac{y_F\sqrt{\rho_\phi}}{\Lambda_1}\right)^{4/11}.
 \end{aligned}
 \label{eq:angle-independent-vacuum-energy}
\end{equation}
Here $\Lambda_1$ is the scale of the running coupling with the Dirac doublet active, $c_E$ is a signed dimensionless coefficient that requires nonperturbative input in the chosen subtraction convention, and $\mathcal E_{\rm ref}$ is an additive vacuum constant. The coefficients $b_1=20/3$ and $b_0=22/3$ give the power $1/11$. Appendix~\ref{app:vacuum-energy} derives this result, states its range of validity and explains the perturbative radial terms included in this convention.

The coefficient $\mathcal A_D(\rho_\phi)$ is positive at the chosen angular minimum and has mass dimension four. It is the periodic amplitude generated by the hidden gauge dynamics. In the same heavy-doublet approximation it is proportional to $\Lambda_D^4(\rho_\phi)$. Its dimensionless coefficient requires a separate nonperturbative determination. We then define the purely angular contribution
\begin{equation}
 \mathcal U_D(\rho_\phi,\vartheta)
 \equiv\mathcal E_D(\rho_\phi,\vartheta)
 -\mathcal E_D(\rho_\phi,0)
 =\mathcal A_D(\rho_\phi)\,[1-\cos\vartheta].
 \label{eq:angular-energy-subtraction}
\end{equation}
Thus $\mathcal U_D$ is the energy cost of changing $\vartheta$ at fixed Higgs magnitude. The angle-independent term $\mathcal E_D(\rho_\phi,0)$ is included in the radial Higgs potential $V_{\rm rad}(\rho_\phi)$, together with its perturbative corrections. Equation~\eqref{eq:radial} gives the leading polynomial approximation to this radial potential near its minimum. With these definitions, the hidden scalar potential is
\begin{equation}
 V_{\rm hid}(\phi,b)=V_{\rm rad}(\rho_\phi)
 +\mathcal U_D(\rho_\phi,\vartheta),
 \qquad
 \rho_\phi=\phi^\dagger\phi,
 \quad
 \vartheta=\theta_D+\arg[\phi e^{-ie_Xb/M_1}].
 \label{eq:covariant-hidden-potential}
\end{equation}
The function $\mathcal U_D$ determines the angular force and the axi-Higgs mass. Describing $\phi$ by an angle requires $|\phi|>0$; passage through $\phi=0$ must be followed with the original scalar and gauge fields.

On a curved spacetime the potential contributes to the action as
\begin{equation}
 S_V=-\int\dd^4x\,\sqrt{-g}\,V_{\rm hid}(\phi,b),
 \qquad g=\det(g_{\mu\nu}).
 \label{eq:covariant-potential-action}
\end{equation}
The factor $\sqrt{-g}$ gives the proper spacetime volume element. The kinetic terms use the gauge covariant derivatives of Eq.~\eqref{eq:derivatives}.

About the broken vacuum, write
\begin{equation}
 \phi=\frac{v_\phi+h}{\sqrt2}e^{i\eta/v_\phi},
 \qquad \langle\phi\rangle=\frac{v_\phi}{\sqrt2}.
 \label{eq:polar}
\end{equation}
The leading masses at this vacuum are
\begin{equation}
 m_h^2=2\lambda_\phi v_\phi^2,\qquad
 m_F=\frac{y_Fv_\phi}{\sqrt2},\qquad
 m_\Psi=\frac{y_\Psi v_\phi}{\sqrt2}.
 \label{eq:radial-expansion-mass}
\end{equation}
The two pseudoscalar fields are $\eta$ and $b$. For real positive $\kappa_{\rm ang}$, the single periodic interaction becomes
\begin{equation}
 -\sqrt2\kappa_{\rm ang}(v_\phi+h)
 \cos\left(\frac{\eta}{v_\phi}-\frac{e_Xb}{M_1}\right).
 \label{eq:single-operator-angle}
\end{equation}
The factor $v_\phi+h$ follows from the assumed linear interaction. For the general gauge potential, its dependence on the Higgs magnitude must be calculated. In our ultralight example the angular energy barely changes the Higgs vacuum value or radial mass, but its magnitude dependence determines the interactions discussed in Section~\ref{sec:gauge-mass}.\\
We choose the coefficient of $H_{\rm SM}^\dagger H_{\rm SM}|\phi|^2$ to vanish at the chosen renormalization scale. Quantum corrections from the heavy portal particles can still affect the visible Higgs mass. Obtaining its observed value therefore requires a protection mechanism or an adjustment of the Higgs parameters.
The different left- and right-handed portal charges can generate gauge anomalies. We calculate their coefficients from fermion loops with three external vertices. The subscript tells which gauge fields meet the loop: $X$ denotes $U(1)_X$ and $D$ denotes $SU(2)_D$.\\
Concerning the resotration of gauge invariance in the hidden sector, the coefficient $C^{(1)}_{XDD}$ describes one portal vector $X_\mu$ and two hidden vectors $W_\nu^a,W_\rho^b$. It is the $U(1)_X[SU(2)_D]^2$ anomaly coefficient of the first model. The hidden-generator trace is proportional to $\delta^{ab}$; we factor out this common tensor when defining $C^{(1)}_{XDD}$. The coefficient $C^{(1)}_{XXX}$ describes three portal-vector vertices. Each vertex contributes the fermion charge, and both hidden-colour components of a doublet contribute to the trace.
The label $X\mathrm{grav}$ denotes one portal-current vertex and two energy--momentum-tensor vertices. These give the mixed gauge--gravitational anomaly. The coefficients below contain the charge sums. Gauge couplings and loop factors enter the corresponding variation of the action.

We evaluate the traces using left-handed Weyl fields. The left-handed charge conjugate of a right-handed fermion of charge $q_R$ has charge $-q_R$. Since both right-handed portal charges in Table~\ref{tab:charges} are $-1$, the same fermion content is represented by
\begin{equation}
 F_L\sim\mathbf2_{+1},\qquad F_R^c\sim\overline{\mathbf2}_{+1},
 \qquad\Psi_L\sim\mathbf1_{+1},\qquad\Psi_R^c\sim\mathbf1_{+1}.
 \label{eq:left-handed-anomaly-fields}
\end{equation}
The conjugate doublet has the same quadratic index as the doublet. Every left-handed entry therefore has portal charge $+1$, with hidden-color multiplicities $2,2,1,1$, respectively.

For the mixed non-Abelian trace, each multiplet contributes its portal charge times its quadratic index. With the generators used above,
\begin{equation}
 \tr_{\mathbf2}(T^aT^b)=\frac14\tr(\sigma^a\sigma^b)
 =\frac12\delta^{ab},\qquad
 T(\mathbf2)=T(\overline{\mathbf2})=\frac12,\qquad T(\mathbf1)=0.
 \label{eq:doublet-index}
\end{equation}
The doublet terms give
\begin{equation}
 C^{(1)}_{XDD}=\sum_{\text{left-handed }f}q_fT(R_f)
 =\underbrace{1\cdot\frac12}_{F_L}
 +\underbrace{1\cdot\frac12}_{F_R^c}
 =(q_{F_L}-q_{F_R})\frac12=1.
 \label{eq:anomalies}
\end{equation}
The factor $1/2$ is the quadratic index obtained by tracing over both doublet components in Eq.~\eqref{eq:doublet-index}. The singlet has $T(\mathbf1)=0$, so the mixed coefficient receives the two displayed doublet contributions.

Three Abelian vertices give the charge factor $q_f^3$. The generators are proportional to the identity within each multiplet, so their trace gives its dimension $d(R_f)$:
\begin{equation}
 \begin{aligned}
 C^{(1)}_{XXX}&=\sum_{\text{left-handed }f}d(R_f)q_f^3\\
 &=\underbrace{2(1)^3}_{F_L}+\underbrace{2(1)^3}_{F_R^c}
 +\underbrace{(1)^3}_{\Psi_L}+\underbrace{(1)^3}_{\Psi_R^c}\\
 &=2\bigl[q_{F_L}^3-q_{F_R}^3\bigr]
 +\bigl[q_{\Psi_L}^3-q_{\Psi_R}^3\bigr]
 =4+2=6.
 \end{aligned}
 \label{eq:cubic-anomaly-trace}
\end{equation}
Here the factor two multiplying the doublet contribution is necessary: an Abelian trace counts both hidden-color components. The right-handed minus sign in the last line is precisely the replacement $q_R\mapsto-q_R$, since $(-q_R)^3=-q_R^3$.

The mixed gravitational triangle has one portal vertex and two energy--momentum-tensor vertices. Its charge sum is linear:
\begin{equation}
 \begin{aligned}
 C^{(1)}_{X\mathrm{grav}}&=\sum_{\text{left-handed }f}d(R_f)q_f\\
 &=\underbrace{2(1)}_{F_L}+\underbrace{2(1)}_{F_R^c}
 +\underbrace{1}_{\Psi_L}+\underbrace{1}_{\Psi_R^c}\\
 &=2\bigl[q_{F_L}-q_{F_R}\bigr]
 +\bigl[q_{\Psi_L}-q_{\Psi_R}\bigr]=4+2=6.
 \end{aligned}
 \label{eq:gravitational-anomaly-trace}
\end{equation}
The hidden fermions generate both the cubic Abelian and mixed gravitational anomalies. The visible contribution cancels with the stated right-handed neutrino content.

The doublet has the same $SU(2)_D$ representation for its two chiralities, but different $U(1)_X$ charges. Its left- and right-handed parts therefore add to the hidden portal anomalies. The two left-handed doublets also satisfy the even-doublet requirement for the global $SU(2)$ anomaly. To cancel the remaining local anomalies, we introduce interactions involving the shifting field $b$ and the topological density $Q_D$ defined in Eq.~\eqref{eq:top-density}.
The coefficients in the action follow from the consistent gauge anomaly, obtained by varying the fermion effective action. For one left-handed Weyl field its cubic Abelian coefficient is $g_X^3q_X^3/(48\pi^2)$ with the definition of the dual field strength in Eq.~\eqref{eq:top-density}. Choose the overall sign so that the total fermion variation is
\begin{equation}
 \delta_\omega\Gamma_{\rm ferm}
 =\int\dd^4x\,\omega\left[
 e_X Q_D+\frac{g_X^3C^{(1)}_{XXX}}{48\pi^2}
 F_X\widetilde F_X+\mathcal A_{\rm grav}\right].
 \label{eq:consistent-anomaly-variation}
\end{equation}
Here $\mathcal A_{\rm grav}$ denotes the mixed gravitational contribution. For the mixed $XDD$ term we assign the anomaly to the Abelian current, preserving the non-Abelian Ward identities. This fixes the finite local counterterm in that triangle amplitude. The cubic term uses the consistent coefficient; the covariant-current coefficient would be three times larger and cannot be substituted in this variation of the action \cite{AnomalyConventions}. Since $\delta b=M_1\omega$, cancellation requires
\begin{equation}
 \mathcal L_{\rm WZ}=-\frac{e_Xb}{M_1}Q_D
 -\frac{b}{4M_1}c_XF_X\widetilde F_X+\mathcal L_{\rm WZ,grav},
 \qquad c_X=\frac{g_X^3}{2\pi^2}.
 \label{eq:wz}
\end{equation}
In particular, the cubic term varies by $-c_X\omega F_X\widetilde F_X/4$, so Eq.~\eqref{eq:consistent-anomaly-variation} gives $c_X=g_X^3C^{(1)}_{XXX}/(12\pi^2)=g_X^3/(2\pi^2)$. The gravitational term cancels the mixed gauge--gravitational anomaly. The Wess--Zumino terms cancel the anomalous gauge variation of the fermion quantum action. Physical fermion-loop amplitudes must still be included. This gives a gauge-consistent effective theory below its cutoff. A theory valid at higher energies that cancels the anomalies through its full particle content has yet to be specified.

The gauge transformations are
\begin{equation}
 X_\mu\mapsto X_\mu+\partial_\mu\omega,
 \qquad b\mapsto b+M_1\omega,
 \qquad\eta\mapsto\eta+e_Xv_\phi\omega.
 \label{eq:transformations}
\end{equation}
The same gauge parameter $\omega$ acts on visible $B-L$ charges and shifts $b$. The separate global visible $B-L$ symmetry protects the Dirac neutrino masses.

At fixed radial magnitude, Eq.~\eqref{eq:lagrangian} gives
\begin{equation}
 \mathcal L_{\rm phase}=\frac12(\partial\eta-e_Xv_\phi X)^2
 +\frac12(\partial b-M_1X)^2.
 \label{eq:phase-action}
\end{equation}
A gauge transformation shifts $(\eta,b)$ in the direction $(e_Xv_\phi,M_1)$. The combination perpendicular to this shift is unchanged and remains physical. With $M_X^2=M_1^2+e_X^2v_\phi^2$, define
\begin{equation}
 G=\frac{e_Xv_\phi\eta+M_1b}{M_X},
 \qquad\chi=\frac{e_Xv_\phi b-M_1\eta}{M_X}.
 \label{eq:rotation}
\end{equation}
These combinations put Eq.~\eqref{eq:phase-action} in the form $\tfrac12(\partial G-M_XX)^2+\tfrac12(\partial\chi)^2$. Since $\delta G=M_X\omega$ and $\delta\chi=0$, $G$ supplies the vector's longitudinal polarization and $\chi$ remains as the physical axion. The gauge-invariant angle and the scale converting it into $\chi$ are
\begin{equation}
 \Theta=\frac{\eta}{v_\phi}-\frac{e_Xb}{M_1}=-\frac{\chi}{F_\chi},
 \qquad F_\chi=\frac{v_\phi M_1}{M_X},
 \qquad F_\chi^{-2}=v_\phi^{-2}+e_X^2/M_1^2.
 \label{eq:F}
\end{equation}
For $M_1\gg e_Xv_\phi$, the phase scale approaches $v_\phi$, with $F_\chi\leq v_\phi$. Putting the Abelian kinetic terms in standard form rescales $g_X$ and $M_1$ together, leaving their ratio in this angle unchanged. We set $\epsilon=0$ at the benchmark renormalization scale; visible fermion loops generate nonzero mixing at other scales.

\section{Tensor origin and scalar realization of the St\"uckelberg field}
\label{sec:stueckelberg-origin}
The action in Eq.~\eqref{eq:lagrangian} specifies the low-energy shifting field $b$ and its anomaly-cancelling interactions. Both a tensor origin and a charged-scalar realization lead to this derivative action. Their high-energy descriptions determine the origin of $M_1$, the fields that stabilize it and their response to inflation. We state these assumptions separately before applying the scalar realization to the numerical example.

\subsection{Tensor and string origin}
\label{sec:tensor-origin}
In string constructions, antisymmetric tensor fields descend from the higher-dimensional theory \cite{OrientifoldAction,StAxions}. In four dimensions, let $C_{\mu\nu}$ be a two-form gauge field with field strength
\begin{equation}
 H_{\mu\nu\rho}=\partial_\mu C_{\nu\rho}
 +\partial_\nu C_{\rho\mu}+\partial_\rho C_{\mu\nu}.
 \label{eq:tensor-strength}
\end{equation}
The kinetic duality follows from the first-order Lagrangian
\begin{equation}
 \mathcal L_{\rm first}=-\frac12J_\mu J^\mu
 +J^\mu(\partial_\mu b-M_1X_\mu)
 -\frac14F_{X\mu\nu}F_X^{\mu\nu}.
 \label{eq:tensor-first-order}
\end{equation}
Eliminating $J_\mu$ gives $\tfrac12(\partial b-M_1X)^2$. Eliminating $b$ instead imposes $\partial_\mu J^\mu=0$, locally solved by $J^\mu=\varepsilon^{\mu\nu\rho\sigma}H_{\nu\rho\sigma}/6$. Integration by parts then converts $-M_1X_\mu J^\mu$ into a $C\wedge F_X$ coupling. The anomaly-cancelling interactions have their corresponding tensor completion \cite{OrientifoldAction,StAxions}.

After canonical normalization, $M_1$ depends on the tensor kinetic coefficient, its vector coupling and the vacuum values of the accompanying moduli. We retain $M_1$ as a parameter of the four-dimensional action; its numerical value requires compactification and stabilization data. The rotation in Eq.~\eqref{eq:rotation} identifies the physical phase after the hidden Higgs acquires its expectation value. Its survival during inflation requires the corresponding moduli to remain stabilized, including the effect of their inflaton couplings.

\subsection{Scalar realization and its symmetry-breaking scale}
\label{sec:scalar-origin}
A charged complex scalar provides an alternative realization of this mass term \cite{ScalarDecoupling}. Let $S$ be an $SU(2)_D$ singlet with portal charge two and write
\begin{equation}
 S=\frac{w+s}{\sqrt2}e^{ib/w},\qquad M_1=e_Xw.
 \label{eq:parent}
\end{equation}
The radial fluctuation $s$ has mass $m_s^2=2\lambda_Sw^2$ for $V_S=\lambda_S(|S|^2-w^2/2)^2$. At energies below this mass, its elimination leaves the phase kinetic term and suppressed higher-derivative interactions. Appendix~\ref{app:s5-scalar-decoupling} gives the derivation and its range of validity. This partial decoupling retains the vacuum value and the angular degree of freedom, as in the Peccei--Quinn construction \cite{PQMechanism}.

With both $S$ and $\phi$ present, the physical field is the gauge-invariant relative phase defined in Eq.~\eqref{eq:F}. An imposed relative-phase symmetry keeps it massless in perturbation theory, and the hidden anomaly generates its small potential. The anomaly-cancelling terms in Eq.~\eqref{eq:wz} accompany either realization of the St\"uckelberg mass.

The scalar vacuum scale is an additional input. For the coupled radial potential in Eq.~\eqref{eq:s5-mag-potential}, stationarity at $I=R=0$ gives
\begin{equation}
 \mu_S^2=\lambda_Sw^2+\frac{\lambda_{\phi S}}{2}v_\phi^2,
 \qquad w=\frac{M_1}{e_X}.
 \label{eq:scalar-scale-input}
\end{equation}
Here $-\mu_S^2$ is the coefficient of $|S|^2$, and $\lambda_{\phi S}$ multiplies $|\phi|^2|S|^2$. The benchmark $M_1=5\times10^{17}\,\GeV$, $e_X=0.6$ gives $w=8.33\times10^{17}\,\GeV$. With the quartics specified in Appendix~\ref{app:s5-scalar-minima}, this corresponds to $\mu_S\simeq5.9\times10^{17}\,\GeV$. The scalar potential therefore specifies a high symmetry-breaking scale whose microscopic origin remains an assumption. The inflationary calculation starts in this broken phase and checks the stability of its minimum.

The coherent-density example uses $F_\chi\simeq10^{17}\,\GeV$, which requires both $v_\phi$ and $w$ to exceed $F_\chi$. The hierarchy $w\gg v_\phi$ makes $F_\chi$ approach $v_\phi$. Equal values $v_\phi=w=\sqrt2F_\chi\simeq1.40\times10^{17}\,\GeV$ also give the same phase normalization, with different vector and fermion thresholds. The numerical results below retain the stated hierarchical benchmark.

The scalar vacuum calculation is developed in Appendix~\ref{app:s5-scalar-minima}, including the displaced minima and radial masses during inflation. The tensor realization of Section~\ref{sec:tensor-origin} requires the analogous calculation for its stabilizing moduli and their couplings.

\section{Protection of the ultralight phase}
\label{sec:phase-protection}
The kinetic terms establish that a physical phase survives. Its small mass also requires control of every interaction that depends on that phase. We use the reduced Planck mass, with $G$ the Newton constant,
\begin{equation}
 \mpl=(8\pi G)^{-1/2}=2.435\times10^{18}\,\GeV.
 \label{eq:reduced-planck}
\end{equation}
In the scalar realization of Section~\ref{sec:stueckelberg-origin}, both $S$ and $\phi$ have portal charge two. The following operator bounds refer to that realization. Their product $S^\dagger\phi$ is gauge invariant. A term
\begin{equation}
 \Delta V=-\left[\mu_{\rm mix}^2 S^\dagger\phi+\mathrm{h.c.}\right]
 =-|\mu_{\rm mix}^2|v_\phi w\cos(\Theta+\delta)
 \label{eq:dangerous-mixing}
\end{equation}
is therefore allowed by the gauge group. The phase $\delta$ is the argument of its coefficient, and $\Theta=-\chi/F_\chi$. This term gives an angular curvature of order $|\mu_{\rm mix}^2|v_\phi w/F_\chi^2$. For a target zero-temperature axi-Higgs mass $m_{\chi,0}\equiv m_\chi(T_D=0)$, the benchmark below requires the gauge susceptibility
\begin{equation}
 \mathcal X_D=m_{\chi,0}^2F_\chi^2\simeq8.9\times10^{-24}\,\GeV^4,
 \qquad \frac{\mathcal X_D}{\mpl^4}\simeq2.5\times10^{-97}.
 \label{eq:quality-size}
\end{equation}
Keeping the extra curvature below one percent of this value requires $|\mu_{\rm mix}^2|\lesssim1.1\times10^{-60}\,\GeV^2$ for a generic phase. An additional symmetry must enforce this suppression.

The imposed relative symmetry forbids such terms in perturbation theory. It is anomalous under the hidden gauge interaction, so the hidden quantum vacuum can still generate the angular energy calculated in the next section. This is the intended source of the axion mass. Other explicit violations of the symmetry, including interactions generated at higher energies, must remain below that contribution.

To quantify the requirement, consider the gauge-invariant series
\begin{equation}
 \Delta V_n=\frac{c_n}{\Lambda_{\rm UV}^{2n-4}}
 (S^\dagger\phi)^n+\mathrm{h.c.},\qquad
 A_n=2|c_n|\Lambda_{\rm UV}^4
 \left(\frac{v_\phi w}{2\Lambda_{\rm UV}^2}\right)^n.
 \label{eq:quality-operators}
\end{equation}
The operator has dimension $2n$, and $A_n$ is its cosine amplitude at the radial vacuum. Its angular frequency is $n/F_\chi$, giving a maximum extra curvature $n^2A_n/F_\chi^2$. A sufficient one-percent condition, without cancellations between unrelated terms, is
\begin{equation}
 n^2A_n<10^{-2}\mathcal X_D.
 \label{eq:quality-bound}
\end{equation}
For illustration take $|c_n|=1$ and $\Lambda_{\rm UV}=\mpl$. The benchmark has $v_\phi w/(2\mpl^2)=0.00703$. The first allowed member of this series must then have $n\geq48$, or dimension at least $96$, with all lower members absent or further suppressed. These numbers depend on the cutoff, coefficients and chosen mass; they measure the severity of the protection required. Interactions outside this series must satisfy the corresponding angular-curvature bound as well.

A discrete gauge symmetry could restrict the allowed powers, but its charge assignments and mixed gauge anomalies must also be consistent. A large discrete subgroup of an anomalous global symmetry cannot simply be assumed to be an exact gauge symmetry with the displayed spectrum. We do not supply the additional fields or anomaly-cancellation mechanism needed for that construction. The present models are conditional on phase protection; their gauge origin determines the surviving field and its couplings. It does not by itself solve this protection problem \cite{GaugedPQ}.

The large scalar scales also limit the effective description. At the benchmark, $v_\phi/\mpl=0.041$, $M_X/\mpl=0.207$ and $w/\mpl=0.342$. The expansion in higher-dimensional scalar interactions and the origin of the anomaly-cancelling terms need information from a theory valid at these energies. The scale $g_X\mpl\simeq7.3\times10^{17}\,\GeV$ suggested by magnetic weak-gravity arguments is comparable to the heavy scales. Such arguments are conjectural and depend on the completion \cite{AxionWGC}; the displayed fields do not establish a controlled gravitational completion.

An axion version of the weak-gravity conjecture concerns an instanton of charge $q$ and action $S_{\rm inst}$ through a condition of order $F_\chi S_{\rm inst}/|q|\lesssim\mpl$ \cite{AxionWGC}. For the small instanton evaluated at $v_\phi$, this ratio is about $16\mpl$ at our one-loop example. The actual potential here is dominated by confinement, where that high-energy semiclassical action is inapplicable. A test of the conjecture requires the relevant instanton spectrum, scale dependence and possible additional states. We neither infer exclusion from the single high-energy action nor claim that this effective theory establishes the conjecture.

\section{Hidden vacuum energy and an ultralight mass}
\label{sec:gauge-mass}
The scalar kinetic terms leave one physical phase. The imposed symmetry keeps it massless in perturbation theory. Hidden gauge dynamics can generate an angular potential; its second derivative at a minimum fixes the axion mass.

The doublet mass contains the phase $\eta/v_\phi$. The chiral field redefinition $F=e^{-i\gamma_5\eta/(2v_\phi)}F'$ makes this mass real. The quantum integration measure then contributes $(\eta/v_\phi)Q_D$. Adding the term in Eq.~\eqref{eq:wz} gives the hidden gauge angle
\begin{equation}
 \vartheta=\theta_D+\frac{\eta}{v_\phi}-\frac{e_Xb}{M_1}
 =\theta_D-\frac{\chi}{F_\chi}.
 \label{eq:theta}
\end{equation}
This change of the hidden gauge angle comes from the doublet $F$. We choose the origin of $\chi$ so that the angular potential has a stable minimum there; the constant $\theta_D$ is then included in that choice.

Fix the Higgs magnitude at $|\phi|=v_\phi/\sqrt2$ and denote the gauge energy by $\mathcal E_D(\vartheta,T_D)$. The hidden temperature $T_D$ is measured in energy units. At nonzero temperature, $\mathcal E_D$ is the free-energy density; at zero temperature it is the ground-state energy density. The second derivative with respect to the angle is called the topological susceptibility. It determines the axion mass:
\begin{equation}
 \mathcal X_D(T_D)=\left.\frac{\partial^2\mathcal E_D}{\partial\vartheta^2}\right|_0,
 \qquad m_\chi^2(T_D)=\frac{\mathcal X_D(T_D)}{F_\chi^2}.
 \label{eq:susceptibility}
\end{equation}
Expanding the energy as $\mathcal E_D=\mathcal E_D(0)+\mathcal X_D\vartheta^2/2+\cdots$ and using Eq.~\eqref{eq:theta} gives the mass relation above. The quantity $\mathcal X_D$ has mass dimension four. For the leading-harmonic form in Eq.~\eqref{eq:hidden-vacuum-energy-form}, its zero-temperature value is $\mathcal X_D(0)=\mathcal A_D(v_\phi^2/2)$. At nonzero temperature the amplitude becomes temperature dependent. Lattice calculations show that the angle dependence changes across the $SU(2)$ thermal transition \cite{SU2Topology}. We keep the cosine as the working approximation in the numerical abundance example.

We can calculate the running gauge coupling, but the overall size of the strong gauge contribution requires separate information. Introduce the one-loop strong scale $\Lambda_0$ below the doublet mass and write
\begin{equation}
 \mathcal X_D(0)=c_0\Lambda_0^4,
 \qquad m_{\chi,0}=\frac{\sqrt{c_0}\Lambda_0^2}{F_\chi}.
 \label{eq:c0}
\end{equation}
The one-loop scale $\Lambda_0$ is a parameter of the running approximation. Its coefficient $c_0$ depends on that convention; setting $c_0=1$ is a choice for the one-loop illustration.

To fix the size of the strong gauge contribution, we use lattice results for pure $SU(2)$ Yang--Mills theory. This theory describes the hidden gauge sector at energies far below $m_F$, where the heavy doublet has been removed; the Higgs and St\"uckelberg fields are $SU(2)_D$ singlets. The lattice calculations evaluate gauge-field observables on a spacetime grid and extrapolate to zero grid spacing. They provide dimensionless ratios that can be applied at the hidden confinement scale.

The common physical reference is the confining string tension $\sigma_D$, the energy per unit length of a long hidden flux tube. We use
\begin{equation}
 \frac{\mathcal X_D(0)^{1/4}}{\sqrt{\sigma_D}}=0.486(10),
 \qquad
 \frac{\Lambda_{\overline{\rm MS}}}{\sqrt{\sigma_D}}=0.752(20)(60).
 \label{eq:lattice-input-ratios}
\end{equation}
The first ratio is the continuum susceptibility estimate in Table~3 of Ref.~\cite{SU2SusceptibilityContinuum}; the conventional $SU(2)$ simulation in Fig.~10 of Ref.~\cite{SU2Susceptibility} is consistent with it. The second ratio is the $SU(2)$ result in Eq.~(26) of Ref.~\cite{SU2Running}, obtained by following the running coupling from finite-volume lattice calculations to the perturbative regime and converting to the $\overline{\rm MS}$ scheme. Its two uncertainties are statistical and estimated systematic errors. The central values give
\begin{equation}
 \mathcal X_D(0)=c_{\overline{\rm MS}}\Lambda_{\overline{\rm MS}}^4,
 \qquad c_{\overline{\rm MS}}=
 \left(\frac{0.486}{0.752}\right)^4\simeq0.17.
 \label{eq:lattice-normalization}
\end{equation}
The uncertainty in the running-scale conversion alone is appreciable, and the quoted determinations are not a joint precision fit. In particular, $c_{\overline{\rm MS}}$ cannot be inserted into a one-loop pole-scale formula while keeping that scale unchanged. At the target physical mass and phase scale, the same lattice inputs imply
\begin{equation}
 \sqrt{\sigma_D}\simeq3.55\,\mathrm{keV},\qquad
 \Lambda_{\overline{\rm MS}}\simeq2.67\,\mathrm{keV},\qquad
 T_c\simeq2.52\,\mathrm{keV}.
 \label{eq:lattice-physical-scales}
\end{equation}
For the last value we used $T_c/\sqrt{\sigma_D}=0.7091(36)$, the continuum $SU(2)$ result in Table~6 of Ref.~\cite{SU2Transition}. This temperature ratio enters the thermal-history calculation. The conversion from the target axion mass to the running scale uses the two ratios in Eq.~\eqref{eq:lattice-input-ratios}:
\begin{equation}
 \mathcal X_D(0)^{1/4}=\sqrt{m_{\chi,0}F_\chi},\qquad
 \Lambda_{\overline{\rm MS}}
 =\frac{0.752}{0.486}\sqrt{m_{\chi,0}F_\chi}.
 \label{eq:lattice-mass-to-scale}
\end{equation}
Thus the chosen mass and phase scale fix the dimensional scale; the lattice calculations supply the conversion factors.

Above $m_F$, the doublet contributes to the running gauge coupling. Below its mass only the $SU(2)$ gauge fields contribute. The one-loop coefficient is $11C_A/3-4T_Fn_f/3$, with $C_A=2$ and $T_F=1/2$. It is therefore $20/3$ for one active Dirac flavor and $22/3$ after that flavor is removed. Joining the running couplings across this mass threshold follows the standard heavy-fermion decoupling procedure \cite{HeavyDecoupling,HeavyMassScaling}; its application to the Higgs-dependent vacuum energy is derived in Appendix~\ref{app:vacuum-energy}. The running equation and its integrated form are
\begin{equation}
 \mu\frac{\dd g_D}{\dd\mu}=-\frac{b_jg_D^3}{16\pi^2},
 \qquad b_1=\frac{20}{3},\quad b_0=\frac{22}{3},
 \qquad
 \alpha_D^{-1}(\mu_2)=\alpha_D^{-1}(\mu_1)
 +\frac{b_j}{2\pi}\ln\!\left(\frac{\mu_2}{\mu_1}\right).
 \label{eq:running}
\end{equation}
The one-loop expression for the coupling diverges at the parameter $\Lambda_0$. To relate this scale to the high-energy coupling, run from $v_\phi$ to $m_F$ with the fermion present, then continue below $m_F$ with the pure-gauge coefficient. Joining the two solutions at $m_F$ gives
\begin{equation}
 \Lambda_0=m_F\exp\left[-\frac1{b_0}
 \left(\frac{2\pi}{\alpha_D(v_\phi)}+b_1\ln\frac{m_F}{v_\phi}\right)\right].
 \label{eq:threshold}
\end{equation}
A prescribed mass therefore requires
\begin{equation}
 \Lambda_0=(m_{\chi,0}F_\chi)^{1/2}c_0^{-1/4},\qquad
 \alpha_D(v_\phi)=\frac{2\pi}
 {b_0\ln(m_F/\Lambda_0)+b_1\ln(v_\phi/m_F)}.
 \label{eq:inverse}
\end{equation}
A perturbative gauge coupling can therefore give a wide range of small masses through its running. Equation~\eqref{eq:inverse} selects a coupling for a prescribed mass. The local sensitivity at fixed $F_\chi,m_F,v_\phi,c_0$ is
\begin{equation}
 \frac{\partial\ln m_{\chi,0}}{\partial\ln\alpha_D(v_\phi)}
 =\frac{4\pi}{b_0\alpha_D(v_\phi)}\simeq104.
 \label{eq:mass-sensitivity}
\end{equation}
A one-percent increase of this coupling changes the one-loop mass by a factor about $2.8$. The many digits of an inverted coupling express numerical inversion within an approximation. They do not specify a comparably accurate physical prediction.

For comparison with Eq.~\eqref{eq:lattice-normalization}, a gauge--fermion two-loop treatment in the $\overline{\rm MS}$ scheme uses
\begin{equation}
 \frac{\dd\alpha_D}{\dd\ln\mu}
 =-\frac{\mathcal B_0(n_f)}{2\pi}\alpha_D^2
  -\frac{\mathcal B_1(n_f)}{8\pi^2}\alpha_D^3,
 \quad \mathcal B_0(n_f)=\frac{22-2n_f}{3},\quad
 \mathcal B_1(n_f)=\frac{136}{3}-\frac{49n_f}{6}.
 \label{eq:two-loop-running}
\end{equation}
Here $n_f$ counts active Dirac doublets. Define
\begin{equation}
 L_{n_f}(\alpha)=\frac{2\pi}{\mathcal B_0\alpha}
 +\frac{\mathcal B_1}{2\mathcal B_0^2}
 \ln\!\left[\frac{\mathcal B_0\alpha/(4\pi)}
 {1+\mathcal B_1\alpha/(4\pi\mathcal B_0)}\right].
 \label{eq:two-loop-scale}
\end{equation}
This is the exact integral of the truncated beta function, with the usual asymptotic definition of the $\overline{\rm MS}$ scale. Setting $\alpha_F=\alpha_D(m_F)$, continuous matching at $\mu=m_F$ gives
\begin{equation}
 L_0(\alpha_F)=\ln\!\left(\frac{m_F}{\Lambda_{\overline{\rm MS}}}\right),
 \qquad L_1(\alpha_D(v_\phi))-L_1(\alpha_F)
 =\ln\!\left(\frac{v_\phi}{m_F}\right).
 \label{eq:two-loop-threshold}
\end{equation}
For the benchmark $m_{\chi,0}=3\times10^{-20}\,\eV$ and $F_\chi=9.93\times10^{16}\,\GeV$, Eq.~\eqref{eq:lattice-mass-to-scale} gives $\mathcal X_D(0)^{1/4}\simeq1.73\,\mathrm{keV}$ and $\Lambda_{\overline{\rm MS}}\simeq2.67\,\mathrm{keV}$. With $v_\phi=10^{17}\,\GeV$ and $y_F=0.2$, the doublet threshold is $m_F=y_Fv_\phi/\sqrt2\simeq1.41\times10^{16}\,\GeV$. The first equation in Eq.~\eqref{eq:two-loop-threshold} then fixes the coupling at this threshold; the second evolves it to $v_\phi$. They give
\begin{equation}
 \alpha_F\simeq0.01648,\qquad
 \alpha_D^{\overline{\rm MS}}(v_\phi)\simeq0.0159.
 \label{eq:lattice-inferred-coupling}
\end{equation}
The high-energy coupling is therefore inferred from the chosen axion mass, the lattice ratios in Refs.~\cite{SU2SusceptibilityContinuum,SU2Running} and the stated two-loop evolution. The lattice calculations themselves determine the pure-gauge ratios. We omit finite higher-order threshold terms and Yukawa and portal contributions to the two-loop evolution. A precision determination would include those terms and propagate the lattice uncertainties.

For orientation, the required potential scale is
\begin{equation}
 \mathcal X_D(0)^{1/4}=31.6\,\eV
 \left(\frac{m_{\chi,0}}{10^{-22}\,\eV}\right)^{1/2}
 \left(\frac{F_\chi}{10^{16}\,\GeV}\right)^{1/2}.
 \label{eq:scale}
\end{equation}
The Higgs expectation value sets the scale between the angle and the field, while the much smaller hidden gauge scale sets the angular potential. The physical phase can thus exist after high-scale symmetry breaking and acquire its relevant potential only when the hidden sector is cold enough.

The Higgs is an $SU(2)_D$ singlet, so its expectation value leaves that gauge group unbroken. The vacuum energy includes long-distance confining effects and large instantons. It must therefore be found from the full strong gauge theory. The factor $\exp[-2\pi/\alpha_D(v_\phi)]$ describes only the small-instanton suppression at high energy. At a fixed high-energy coupling, joining the running solutions gives
\begin{equation}
 \Lambda_0(h)^{b_0}=m_F(h)^{b_0-b_1}\Lambda_1^{b_1},
 \qquad m_F(h)=\frac{y_F(v_\phi+h)}{\sqrt2},
 \qquad\mathcal X_D(h)\propto(v_\phi+h)^{4/11}
 \label{eq:radial-threshold}
\end{equation}
when the fermion is much heavier than the confinement scale and the pure-gauge vacuum-energy coefficient is held fixed. Here $\Lambda_1$ is the scale defined with one active Dirac flavour. This formula also gives the dependence on the Higgs fluctuation $h$ defined in Eq.~\eqref{eq:polar}. Its fractional power differs from the linear factor $v_\phi+h$ in Eq.~\eqref{eq:single-operator-angle}. At the vacuum we can choose a cosine coefficient that gives the same axion mass:
\begin{equation}
 \lambda_{\rm eff}=\frac{\mathcal X_D(0)}{v_\phi^4}.
 \label{eq:lambdaeff}
\end{equation}
This choice makes the second angular derivative agree at the radial vacuum. Calculating radial decays also requires how the potential and kinetic terms change with the Higgs magnitude. The different magnitude dependences in Eqs.~\eqref{eq:single-operator-angle} and \eqref{eq:radial-threshold} therefore lead to different radial couplings.

\begin{figure}[tb]
\centering
\begin{tikzpicture}
\begin{semilogyaxis}[width=.65\textwidth,height=4.8cm,
 xlabel={$\alpha_D(v_\phi)$},ylabel={$m_{\chi,0}\,[\mathrm{eV}]$},
 grid=major,xmin=.014,xmax=.018,scaled x ticks=false,
 legend style={font=\scriptsize,at={(.5,-.32)},anchor=north,draw=none},
 xtick={.014,.015,.016,.017,.018},
 xticklabel style={/pgf/number format/fixed,/pgf/number format/precision=3}]
\addplot[blue!70!black,thick] table[x=alpha,y=mass_eV]{mass_scan.dat};
\addlegendentry{One loop, $c_0=1$}
\addplot[orange!80!black,dashed,thick] table[x=alpha,y=mass_eV]{mass_scan_lattice.dat};
\addlegendentry{Two loop, lattice ratios}
\addplot[only marks,mark=*,red!70!black,forget plot] coordinates {(.01640247218,3e-20)};
\addplot[only marks,mark=triangle*,orange!80!black,forget plot] coordinates {(.01593361765,3e-20)};
\end{semilogyaxis}
\end{tikzpicture}
\caption{Mass as a function of the selected high-energy gauge coupling for $v_\phi=10^{17}\,\GeV$, $y_F=0.2$, $M_1=5\times10^{17}\,\GeV$ and $g_X=0.3$. The solid curve uses one-loop running and $c_0=1$. The dashed curve uses the gauge--fermion two-loop beta function and the central pure-$SU(2)$ ratios in Eq.~\eqref{eq:lattice-input-ratios}, from Refs.~\cite{SU2SusceptibilityContinuum,SU2Running}. Both curves can reach the chosen mass $3\times10^{-20}\,\eV$, at the marked couplings. Their displacement illustrates the scale-conversion and running dependence. Neither curve assigns a preferred axion mass, and the dashed curve omits the additional uncertainties stated in the text.}
\label{fig:mass}
\end{figure}
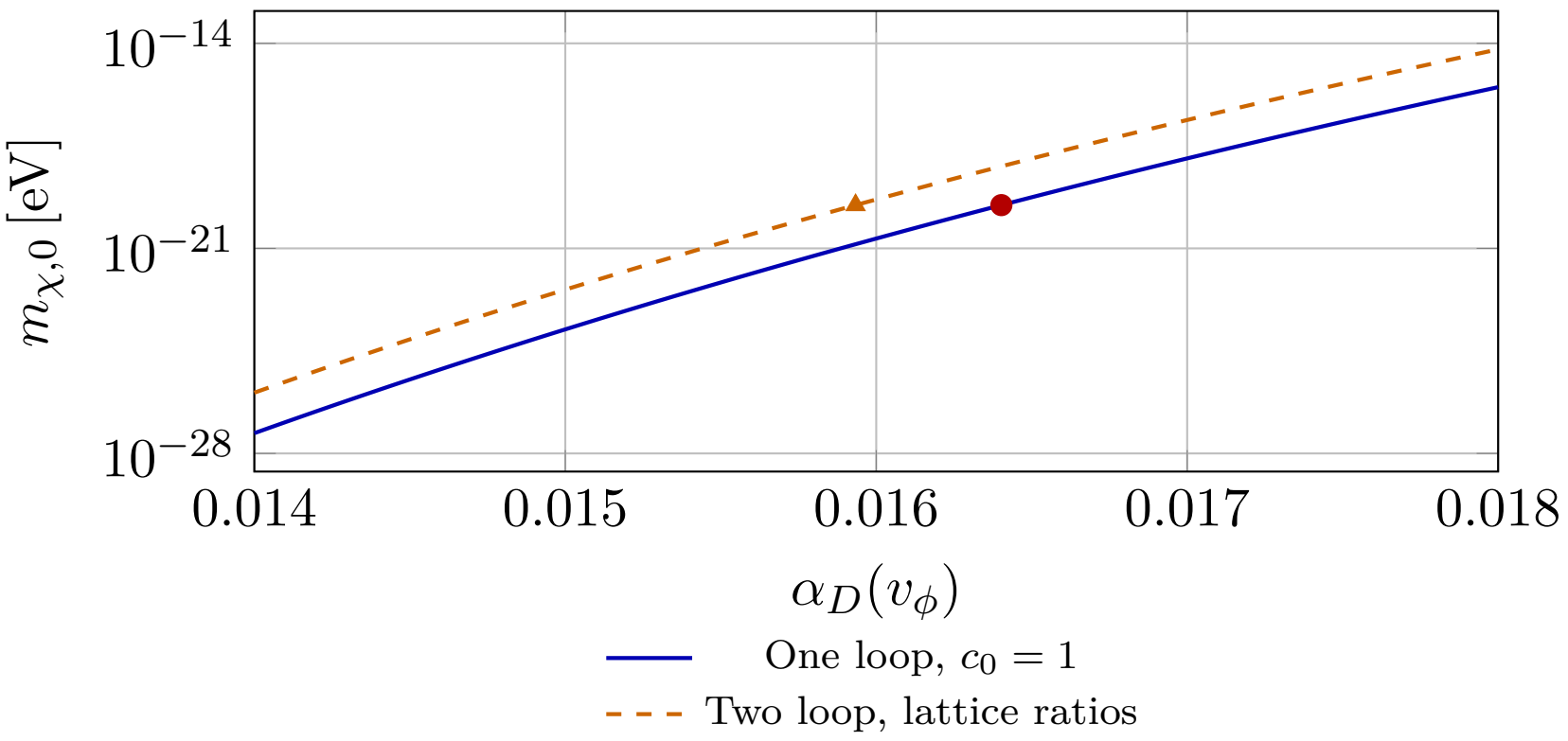

\section{Inflation and vacuum stability}
\label{sec:inflation-embedding}
The relic-density calculation starts from a nearly uniform field that survives inflation and reheating. We give a conditional example using the scalar realization of Section~\ref{sec:scalar-origin}. The inflaton $I$ is a separate real scalar, neutral under the gauge groups, with a canonical kinetic term and the plateau potential of Ref.~\cite{PolynomialInflation},
\begin{equation}
 S_I=\int\dd^4x\sqrt{-g}\left[
 \frac12g^{\mu\nu}\partial_\mu I\partial_\nu I-U(I)\right],
 \qquad U(I)=U_0\frac{I^2}{I^2+\mu_I^2}.
 \label{eq:inflaton-potential}
\end{equation}
The background evolution, primordial spectrum and matching to reheating are derived in Appendix~\ref{app:s5-inflation}. The scalar potential and its couplings to $I$ and spacetime curvature are specified in Appendix~\ref{app:s5-scalar-minima}. These interactions preserve the relative-phase symmetry. The phase protection and the small hidden gauge mass during inflation remain assumptions of the example.

For visible reheating at $T_R=10^{10}\,\GeV$, the amplitude and expansion conditions give
\begin{equation}
 \begin{gathered}
 \mu_I\simeq6.05\times10^{15}\,\GeV,\qquad
 U_0\simeq1.13\times10^{61}\,\GeV^4,\qquad
 H_*\simeq7.96\times10^{11}\,\GeV,\\
 N_*\simeq49.8,\qquad
 \mathcal P_{\mathcal R}(k_*)\simeq2.10\times10^{-9},\qquad
 n_s\simeq0.9694,\qquad r\simeq1.03\times10^{-5}.
 \end{gathered}
 \label{eq:inflation-example}
\end{equation}
Here $k_*=0.05\,\mathrm{Mpc}^{-1}$, and $N_*$ counts expansion to the first crossing of $\epsilon_H=1$. The subsequent oscillatory expansion is included in the reheating matching. The spectrum uses first-order Hubble slow-roll expressions. The comparison with the published scalar-spectrum constraints and the nearby examples are retained in Appendix~\ref{app:s5-isocurvature} \cite{Inflation,ACTBase,ACTInflation}.

The specified inflaton and curvature couplings change $v_\phi^2$ by about $-7.6\times10^{-11}$ of its vacuum value and $w^2$ by about $-2.1\times10^{-14}$. Both radial eigenmasses remain much larger than $H_*$, and the fractional change of $F_\chi$ is about $-3.7\times10^{-11}$. The displaced vacuum and its tree-level effect on the inflaton potential are calculated in Appendix~\ref{app:s5-scalar-minima}; Appendix~\ref{app:s5-loops} gives the heavy-particle loop estimates. These results support the use of the fixed phase scale in this scalar example. A tensor realization requires its own moduli-stabilization calculation.

For the coherent dark-matter benchmark, the angular fluctuations give $\mathcal P_S\simeq6.3\times10^{-11}$ and $\beta_{\rm iso}\simeq0.029$ under the uncorrelated-isocurvature assumptions. The corresponding amplitude bound is $H_*\lesssim9.2\times10^{11}\,\GeV$ \cite{Inflation}. Appendix~\ref{app:s5-isocurvature} derives the density response and this bound. A joint likelihood would fit the scalar and isocurvature spectra together.

The visible decay interaction and its perturbative reheating conditions are developed in Appendix~\ref{app:s5-reheating}. The energy bound on the maximum temperature, calculated in Appendix~\ref{app:s5-temperature}, leaves both scalar vacuum values nonzero under the stated positive thermal-mass bounds. Inflaton self-resonance can change the late pressure history. Replacing the late averaged pressure by a radiationlike stage gives $N_*\simeq53.3$ and $n_s\simeq0.9714$ after repeating the matching; a nonlinear fragmentation calculation is needed to determine the actual pressure.

The visible decay operator also changes the heavy-vector mass through hypercharge--portal kinetic mixing. Appendix~\ref{app:s5-kinetic-mixing} finds an uncancelled slope correction of order $3\times10^2$ times the required slope for $\epsilon=10^{-3}$. Suppressing that contribution below one percent would require $|\epsilon|\lesssim6\times10^{-6}$ at the matching scale. Visible fermions generate mixing through their hypercharge and $B-L$ charges. The chosen plateau is therefore a renormalized potential whose coefficients require adjustment or protection at high energies.

The cold hidden history further requires $B_D/B_{\rm vis}\simeq4.6\times10^{-12}$. The relation between this energy fraction, the hidden temperature normalization and the inflaton decay coefficients is derived in Appendix~\ref{app:s5-temperature}. The displayed symmetries permit both visible and hidden decay operators, so the small hidden coefficient needs a microscopic explanation. The vacuum and isocurvature results apply to the specified scalar potential, phase protection and reheating history.

\section{Thermal history and initial field values}
\label{sec:history}
The relic density depends on when the axion starts to move. This is determined by the expansion rate and the axion mass at that time. A hot hidden plasma can reduce the angular potential and delay oscillations compared with an estimate using the zero-temperature mass.

Section~\ref{sec:inflation-embedding} gives one inflationary example in which the hidden Higgs vacuum survives. Other inflaton potentials and couplings must pass the same vacuum and isocurvature tests. We use the surviving Higgs values and phase as the starting point for the later thermal evolution.

We assume that both Higgs magnitudes are already nonzero, with a nearly common phase across the relevant region, before observable modes leave the inflationary horizon. Their vacuum values must remain nonzero during reheating, including its hottest stage. We also take the radial masses to be much larger than the inflationary Hubble rate. The late initial field values are then
\begin{equation}
 \chi_i=F_\chi\theta_i,\qquad \dot\chi_i\simeq0.
 \label{eq:initial}
\end{equation}
The initial angle $\theta_i$ measures the displacement from the angular minimum. We choose it as an initial condition. Enough inflation can stretch a region with this angle to contain our observable Universe. If the physical phase forms after inflation, different regions generally have different initial angles, and the density calculation must include their spatial variation and any defects allowed by the model.

The visible and hidden sectors can have different temperatures. While the six hidden gluon degrees of freedom are deconfined and in thermal equilibrium among themselves, define $T_D/T=\xi(T)$. If each sector conserves its own entropy during expansion,
\begin{equation}
 \xi(T)=\xi_*\left[\frac{g_{*s}(T)}{g_{*s,*}}\right]^{1/3},
 \qquad g_{*s,*}=\frac{43}{11},
 \label{eq:xi}
\end{equation}
A direct measure of the hidden population is its entropy relative to visible entropy,
\begin{equation}
 \mathcal R_D\equiv\frac{s_D}{s_{\rm ref}}
 =\frac{6}{43/11}\,\xi_*^3.
 \label{eq:hidden-entropy-ratio}
\end{equation}
The parameter $\xi_*$ is a convenient way to specify this ratio. It equals a temperature ratio only when the six hidden gluon degrees of freedom remain relativistic and the visible entropy count is $43/11$. For the benchmark, $\mathcal R_D=1.53\times10^{-9}$. Equation~\eqref{eq:xi} gives the temperature ratio during the deconfined era. We conserve $\mathcal R_D$ through an internally equilibrated transition with negligible transfer to other fields. If the transition produces additional entropy, its post-transition value must instead be supplied.

The physical transition temperature can be taken from Eq.~\eqref{eq:lattice-physical-scales}. Its relation to a running scale depends on the scale convention. A constant-mass approximation requires small changes in the angular curvature at the actual oscillation time. In thermal equilibrium write
\begin{equation}
 \mathcal X_D(T_D)=\mathcal X_D(0)\,\mathcal F(T_D/T_c),
 \qquad\mathcal F(0)=1.
 \label{eq:thermal-mass}
\end{equation}
If oscillations start near this transition, the temperature dependence of the susceptibility must be calculated or obtained from strong-coupling data.

The confined population consists of massive glueballs. A gas description well below the transition is supported by lattice calculations of the confined equation of state \cite{SU2EquationOfState}. Near the transition its full spectrum and equation of state are needed. To explain the later evolution, let $G$ denote a light scalar glueball of mass $m_G$ and degeneracy one. Let $n_G,T_D,\rho_G,P_G$ denote its number density, kinetic temperature, energy density and pressure. In a dilute nonrelativistic gas,
\begin{equation}
 n_G^{\rm eq}(T_D)=\left(\frac{m_GT_D}{2\pi}\right)^{3/2}e^{-m_G/T_D},
 \quad \rho_G\simeq n_G\left(m_G+\frac32T_D\right),\quad P_G=n_GT_D.
 \label{eq:glueball-gas}
\end{equation}
Elastic collisions redistribute momentum. Reactions with three incoming and two outgoing glueballs also reduce their number, converting rest energy into kinetic energy. With the reaction coefficient defined to include identical-particle factors, a one-species treatment is
\begin{equation}
 \dot n_G+3Hn_G=-\langle\sigma_{3\to2}v^2\rangle
 \left(n_G^3-n_G^2n_G^{\rm eq}\right),\qquad
 \dot\rho_G+3H(\rho_G+P_G)=0.
 \label{eq:glueball-evolution}
\end{equation}
These equations neglect energy transfer to the axion and visible sectors, and assume that elastic scattering maintains a momentum distribution described by $T_D$. Other populated glueballs require coupled number equations \cite{Glueballs}.

When the number-changing rate $\Gamma_{32}=n_G^2\langle\sigma_{3\to2}v^2\rangle$ is much larger than $H$, the chemical potential is approximately zero and $n_G\simeq n_G^{\rm eq}$. Writing $x_G=m_G/T_D$, the entropy is then
\begin{equation}
 s_D\simeq n_G^{\rm eq}(x_G+5/2),\qquad
 x_G+\frac12\ln x_G=3\ln(a/a_\dagger)+\mathrm{constant}
 \quad(x_G\gg1).
 \label{eq:cannibal-temperature}
\end{equation}
The scale factor $a_\dagger$ is any fixed reference. The second relation follows by inserting Eq.~\eqref{eq:glueball-gas} in $s_Da^3=\mathrm{constant}$ and keeping the leading large-$x_G$ terms. The hidden gas cools approximately as $1/\ln a$, because number-changing reactions keep supplying kinetic energy. This is often called a cannibal phase. After these reactions cease, $n_Ga^3$ stays nearly fixed and a freely redshifting nonrelativistic momentum distribution gives $T_D\propto a^{-2}$. If chemical equilibrium ends at $x_f$, the later number-to-visible-entropy ratio is approximately
\begin{equation}
 Y_G\equiv\frac{n_G}{s_{\rm ref}}\simeq\frac{\mathcal R_D}{x_f+5/2}.
 \label{eq:glueball-yield}
\end{equation}
The freeze-out value $x_f$ must be found from the reaction rate and the expansion. Continuing $T_D\propto a^{-1}$ through confinement describes neither of these nonrelativistic limits.

At fixed $F_\chi$, the homogeneous equation and the onset estimate are
\begin{equation}
 \ddot\chi+3H\dot\chi+
 \frac{\dd V(\chi,T_D)}{\dd\chi}=0,
 \qquad 3H(t_{\rm osc})\sim m_\chi(T_D(t_{\rm osc})).
 \label{eq:evolution}
\end{equation}
The term $\ddot\chi$ is the time acceleration of the field value. The expansion term $3H\dot\chi$ damps its motion when $H\gg m_\chi$. In our cosine example, $V=\mathcal X_D(T_D)[1-\cos(\chi/F_\chi)]$. The condition $3H\sim m_\chi$ estimates the onset; the differential equation follows the gradual development of the oscillations.

The two sectors exchange energy efficiently when the portal interaction rate exceeds the expansion rate. Well below $M_X$, exchange of the heavy vector gives a dimension-six current interaction and the estimate
\begin{equation}
 \begin{aligned}
 \Gamma_{\rm portal}\sim C\frac{g_X^4T^5}{M_X^4},
 &\qquad H=\sqrt{\frac{\pi^2g_*}{90}}\frac{T^2}{\mpl}.
 \end{aligned}
 \label{eq:contact}
\end{equation}
The coefficient $C$ contains the charges and phase-space factors. The estimate applies when the final particles can be produced at the available energy. Heavy hidden fermions further reduce production. A small rate lets the two sectors keep different temperatures after reheating, although the initial temperature difference must be specified separately. Visible loops also generate kinetic mixing; exchange through the heavy vector remains suppressed in this history.

\section{Coherent axion density during radiation domination}
\label{sec:abundance}
The coherent component is the nearly uniform field oscillating with a common phase across the region considered. A quantum state with a large coherent occupation and a nonzero field expectation value can be followed with the classical field equation. We count its energy in this mean field. Additional quanta outside that coherent component are described by a separate momentum distribution.

Suppose the mass has settled before oscillations and the angle is small enough for a quadratic potential. Radiation domination gives $H=1/(2t)$ when the relativistic degrees of freedom are constant. With $x=m_{\chi,0}t$, the field equation becomes
\begin{equation}
 \frac{\dd^2\chi}{\dd x^2}+\frac{3}{2x}\frac{\dd\chi}{\dd x}+\chi=0.
 \label{eq:dimensionless}
\end{equation}
Set $\chi=x^{-1/4}u$. Direct substitution gives
\begin{equation}
 u''+\frac1x u'+\left(1-\frac{1}{16x^2}\right)u=0.
 \label{eq:bessel}
\end{equation}
The regular solution is a Bessel function of order $1/4$. Its constant multiplier is fixed by $\chi\to\chi_i$ as $x\to0$:
\begin{equation}
 \chi(x)=\chi_i\,2^{1/4}\Gamma(5/4)x^{-1/4}J_{1/4}(x).
 \label{eq:solution}
\end{equation}
Indeed $J_{1/4}(x)\simeq(x/2)^{1/4}/\Gamma(5/4)$ at small $x$. At large $x$, its oscillatory amplitude is proportional to $x^{-3/4}$. Squaring the amplitude and adding the averaged kinetic and potential energies yields
\begin{equation}
 \langle\rho_{\rm coh}\rangle
 =C_R\chi_i^2m_{\chi,0}^2x^{-3/2},
 \qquad C_R=\frac{\sqrt2\,\Gamma(5/4)^2}{\pi}=0.369834\ldots.
 \label{eq:energy}
\end{equation}
During radiation domination $a\propto t^{1/2}$, so $x^{-3/2}\propto a^{-3}$. The averaged energy density of the oscillating field therefore decreases in the same way as that of nonrelativistic matter.

To compare densities at different times, use the entropy density of the visible bath,
\begin{equation}
 s_{\rm ref}=B_sT^3,\qquad B_s=\frac{2\pi^2}{45}g_{*s},
 \qquad A_R=\sqrt{\frac{\pi^2g_*}{90}},\qquad
 T^2=\frac{m_{\chi,0}\mpl}{2A_Rx}.
 \label{eq:entropy}
\end{equation}
If the visible bath conserves entropy in a region expanding with the Universe, its entropy density decreases as $a^{-3}$. Once the axion has a constant mass and oscillates rapidly, its averaged energy density decreases by the same factor. Their ratio then stays constant. Inserting Eq.~\eqref{eq:entropy} into Eq.~\eqref{eq:energy} cancels the time variable $x$:
\begin{equation}
 \frac{\rho_{\rm coh}}{s_{\rm ref}}
 =K_R\frac{F_\chi^2\theta_i^2\sqrt{m_{\chi,0}}}{\mpl^{3/2}},
 \qquad K_R=C_R\frac{(2A_R)^{3/2}}{B_s}.
 \label{eq:relic}
\end{equation}
After electron--positron annihilation, $g_*=3.36$ and $g_{*s}=43/11$, giving $K_R=0.288507$. The entropy count includes photons and the standard relic neutrinos. If visible comoving entropy later grows by a factor $D$, divide Eq.~\eqref{eq:relic} by $D$. A changing axion mass generally requires the full field equation. Approximate conservation of the oscillation action can be used when the field oscillates rapidly and the mass changes little during each period.

Using the reference values $s_0=2891.2\,\mathrm{cm}^{-3}$ and $\rho_c/h_0^2=1.05375\times10^{-5}\,\GeV\,\mathrm{cm}^{-3}$, the density parameter is
\begin{equation}
 \Omega_{\rm coh}h_0^2
 =\frac{s_0}{\rho_c/h_0^2}\frac{\rho_{\rm coh}}{s_{\rm ref}},
 \qquad \left(\frac{\rho_{\rm DM}}{s_{\rm ref}}\right)_{\rm ref}
 =4.37362\times10^{-10}\,\GeV
 \label{eq:omega}
\end{equation}
for the reference dark-matter value $\Omega_{\rm DM}h_0^2=0.12$. Here $h_0$ is the present Hubble constant in units of $100\,\mathrm{km\,s^{-1}\,Mpc^{-1}}$. These values set the density to be reproduced. Testing the resulting matter distribution requires a later perturbation calculation.

For angles beyond the quadratic approximation, the shape of the potential must be specified. For a cosine we solve
\begin{equation}
 \theta''+\frac{3}{2x}\theta'+\sin\theta=0,
 \qquad \theta(0)=\theta_i,\quad\theta'(0)=0.
 \label{eq:cosine}
\end{equation}
Once the oscillations are rapid, the following averaged dimensionless quantity approaches a constant:
\begin{equation}
 I(\theta_i)=\left\langle x^{3/2}
 \left[\frac12\theta'^2+1-\cos\theta\right]\right\rangle,
 \qquad
 \frac{\rho_{\rm coh}}{s_{\rm ref}}
 =\frac{(2A_R)^{3/2}}{B_s}
 \frac{F_\chi^2\sqrt{m_{\chi,0}}}{\mpl^{3/2}}I(\theta_i).
 \label{eq:invariant}
\end{equation}
In the quadratic limit this constant is $I=C_R\theta_i^2$. For larger initial angles, we evolve Eq.~\eqref{eq:cosine} from the initially slow-moving field until the average in Eq.~\eqref{eq:invariant} stops changing. This gives the correction to the quadratic relic density. For still larger angles, a reliable prediction requires the angular shape from the full strong gauge theory.

\section{An illustrative high-scale example}
\label{sec:benchmark}
We select a mass, phase scale and initial displacement to illustrate the density relations. The mass is reached by choosing the high-energy coupling; the angle is then chosen to give the reference abundance. The hidden entropy and phase protection are separate inputs. This procedure establishes a conditional parameter example and its required scales.

Take
\begin{equation}
 \begin{gathered}
 v_\phi=10^{17}\,\GeV,\qquad M_1=5\times10^{17}\,\GeV,
 \qquad g_X=0.3,\\
 \lambda_\phi=0.01,\qquad y_F=y_\Psi=0.2,
 \qquad m_{\chi,0}=3\times10^{-20}\,\eV,\qquad c_0=1.
 \end{gathered}
 \label{eq:benchmark-input}
\end{equation}
The preceding equations give the phase scale and heavy-particle masses:
\begin{equation}
 \begin{gathered}
 M_X=5.036\times10^{17}\,\GeV,\qquad
 F_\chi=9.929\times10^{16}\,\GeV,\\
 m_h=m_F=m_\Psi=1.414\times10^{16}\,\GeV,
 \qquad w=8.333\times10^{17}\,\GeV.
 \end{gathered}
 \label{eq:benchmark-heavy}
\end{equation}
We take the radial mass $m_\rho$ above the vector and Higgs masses, so its effects at later, lower energies are small. All stated vacuum values lie below $\mpl$. Explaining these different scales and protecting the visible Higgs mass require information about the theory at higher energies.

Equation~\eqref{eq:inverse} gives
\begin{equation}
 \Lambda_0=1.726\,\mathrm{keV},\qquad
 \alpha_D(v_\phi)=0.01640,\qquad g_D(v_\phi)=0.4540.
 \label{eq:benchmark-gauge}
\end{equation}
Using the pure-$SU(2)$ ratios in Refs.~\cite{SU2SusceptibilityContinuum,SU2Running} and two-loop running gives $\alpha_D^{\overline{\rm MS}}(v_\phi)\simeq0.0159$ for the same physical target, as derived in Eqs.~\eqref{eq:lattice-mass-to-scale}--\eqref{eq:lattice-inferred-coupling}. All subsequent homogeneous density results depend on the selected physical mass and phase scale. In the one-loop illustration the coupling gives $\exp[-2\pi/\alpha_D(v_\phi)]\simeq4.34\times10^{-167}$. The axion mass from confinement corresponds to $\lambda_{\rm eff}=8.87\times10^{-92}$ in Eq.~\eqref{eq:lambdaeff}. The first number is the high-energy small-instanton suppression; the second reproduces the angular potential from confinement. Changing $c_0$ changes the required confinement scale by $c_0^{-1/4}$. An accurate high-energy coupling requires a strong-gauge calculation of this coefficient using the same definition of the confinement scale.

Choose $\xi_*=10^{-3}$, equivalently $\mathcal R_D=1.53\times10^{-9}$. The physical transition scale is $T_c\simeq2.52\,\mathrm{keV}$ from the lattice ratios. With the constant mass, the visible temperature at the conventional onset estimate is
\begin{equation}
 T_{\rm osc}\simeq6.33\,\mathrm{keV}.
 \label{eq:benchmark-onset}
\end{equation}
The hidden temperature must be found from the confined gas. As a one-species illustration, take the rounded mass $m_G=4\sqrt{\sigma_D}\simeq14.2\,\mathrm{keV}$. This is a representative input for the illustration. If number-changing reactions are still fast, solve
\begin{equation}
 \left(\frac{m_GT_D}{2\pi}\right)^{3/2}e^{-m_G/T_D}
 \left(\frac{m_G}{T_D}+\frac52\right)
 =\mathcal R_Ds_{\rm ref}(T_{\rm osc}).
 \label{eq:confined-onset-temperature}
\end{equation}
It gives $x_G\simeq18.1$ and $T_D\simeq0.78\,\mathrm{keV}$, with visible photons still at $6.33\,\mathrm{keV}$. Parametrize the unknown reaction coefficient by $\langle\sigma_{3\to2}v^2\rangle=c_{32}/m_G^5$. The corresponding rate is $\Gamma_{32}/H\simeq182c_{32}$; this chemical-equilibrium estimate requires $182c_{32}\gg1$. Earlier freeze-out gives a different density and temperature and is treated with Eq.~\eqref{eq:glueball-evolution}. The small finite-density contribution to the angular potential can be bounded independently of this rate, as shown in Section~\ref{sec:consistency}.

The quadratic abundance gives an initial angle $\theta_i=0.32662$. Direct evolution of the illustrative cosine gives
\begin{equation}
 \theta_i=0.32520,\qquad
 \chi_i=3.23\times10^{16}\,\GeV,\qquad
 \Omega_{\rm coh}h_0^2\simeq0.1200.
 \label{eq:benchmark-abundance}
\end{equation}
Using the cosine changes the required initial angle by less than one percent. We checked numerical convergence and agreement with the analytic quadratic solution. The main physical uncertainties come from the hidden potential and thermal history. The coherent field gives the reference dark-matter density; a small additional population can be included by slightly reducing $\theta_i$.

\begin{figure}[tb]
\centering
\begin{tikzpicture}
\begin{axis}[width=.65\textwidth,height=4.8cm,
 xlabel={$\theta_i$},ylabel={$\rho_{\rm coh}/\rho_{\rm DM,ref}$},
 grid=major,xmin=0,xmax=1,ymin=0,legend pos=north west]
\addplot[blue!70!black,thick] table[x=theta,y=fraction]{abundance_scan.dat};
\addlegendentry{Illustrative cosine}
\addplot[red!65!black,dashed,thick,domain=.01:1,samples=80]{(x/.326618322)^2};
\addlegendentry{Quadratic limit}
\addplot[only marks,mark=*,black] coordinates {(.3252048135,1)};
\end{axis}
\end{tikzpicture}
\caption{Coherent axion density for the mass and phase scale in Eqs.~\eqref{eq:benchmark-input}--\eqref{eq:benchmark-heavy}. The solid curve shows the numerical cosine example; the dashed curve gives the analytic quadratic approximation. The marked initial angle reproduces the reference dark-matter density. Values above unity give too much dark matter for the assumed history.}
\label{fig:abundance}
\end{figure}
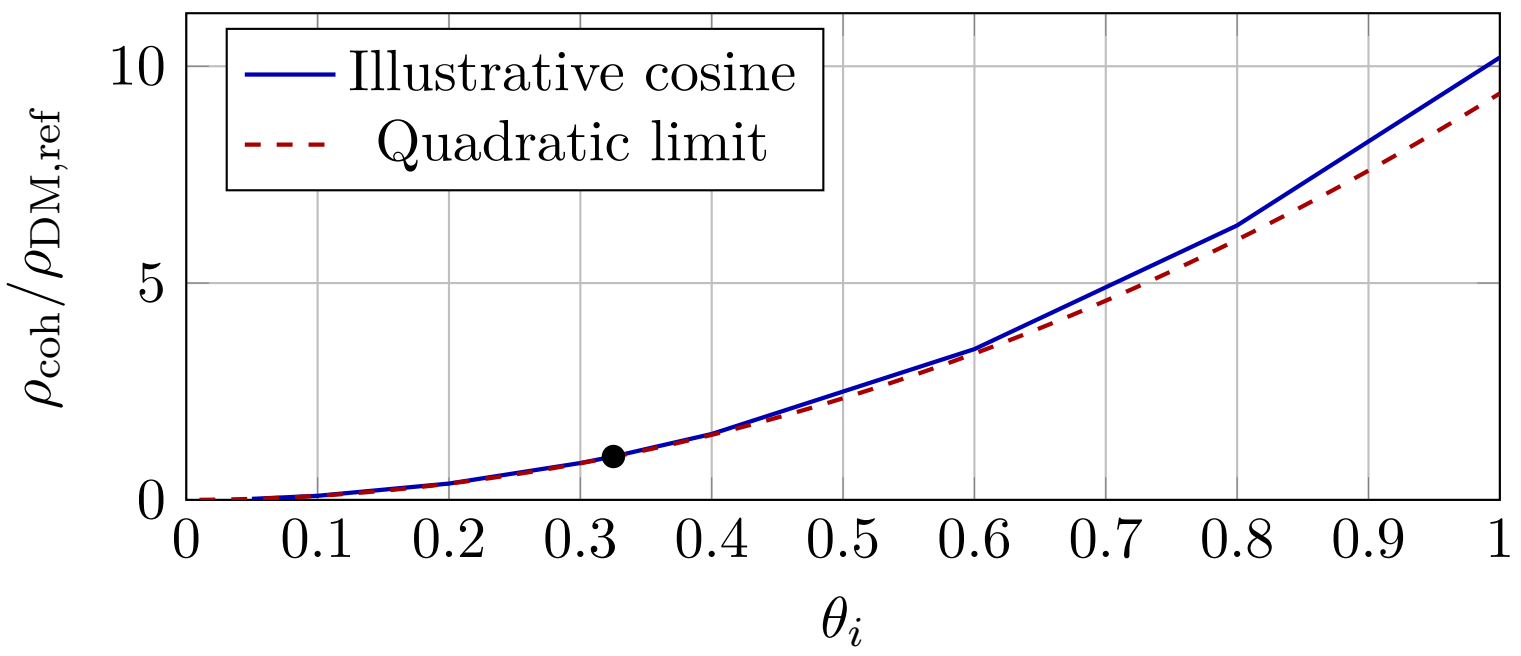

\section{Other populations and axion survival}
\label{sec:consistency}
The dark sector can contain the coherent axion, hidden gauge radiation, glueballs and other particles produced earlier. Inflation also produces fluctuations of the axion field. We give the conditions under which these effects are compatible with the coherent dark-matter example.

For $T_R=10^{10}\,\GeV$, Eq.~\eqref{eq:contact} gives $\Gamma_{\rm portal}/H\sim9\times10^{-26}C$. The charged hidden fermions are much heavier than the bath temperature, which further reduces their production. We assume that the earlier hottest stage preserves the Higgs vacuum and produces negligible heavy-particle populations. The small hidden entropy is chosen through the reheating condition. If reheating produces appreciable numbers of $\Psi$, $F$ or radial particles, their distribution functions must also be evolved.

The small confined contribution can be bounded without choosing a $3\to2$ coefficient. At an equilibrated transition with negligible chemical potentials,
\begin{equation}
 \frac{\rho_{D,c}}{s_{\rm ref,c}}
 =\mathcal R_D\frac{\rho_{D,c}}{s_{D,c}}
 =\mathcal R_D T_c\frac{\rho_{D,c}}{\rho_{D,c}+P_{D,c}}
 \leq\mathcal R_D T_c.
 \label{eq:glueball-energy}
\end{equation}
The pressure and energy here are the thermal contributions with the zero-temperature vacuum part subtracted. We assume nonnegative pressure and a common equilibrium transition. Subsequently, negligible energy exchange and $P_G\geq0$ imply
\begin{equation}
 \frac{\dd(\rho_Ga^3)}{\dd t}=-3Ha^3P_G\leq0,
 \qquad\frac{\rho_G}{s_{\rm ref}}\leq\mathcal R_DT_c
 \simeq3.9\times10^{-6}\,\eV.
 \label{eq:glueball-late-bound}
\end{equation}
This is at most $8.9\times10^{-6}$ of the reference dark-matter energy per unit entropy. The bound includes the pressure work during a cannibal phase and remains valid after number-changing freeze-out. It assumes the stated entropy history and excludes additional production or energy injection. In the chemical-equilibrium example at onset, Eq.~\eqref{eq:glueball-gas} instead gives $\rho_G/s_{\rm ref}\simeq1.1\times10^{-6}\,\eV$. A precise final glueball abundance requires its full spectrum and reaction coefficients \cite{Glueballs}.

For a dilute confined population, its leading effect on a slowly varying axion is the angular dependence of the glueball energies. At a CP-even minimum, $m_G'(0)=0$, and a nonrelativistic population contributes approximately
\begin{equation}
 \begin{aligned}
 \Delta\mathcal X_D&\simeq n_Gm_G''(0),\qquad
 C_G\equiv\frac{|m_G''(0)|}{m_G},\\
 \frac{|\Delta\mathcal X_D|}{\mathcal X_D}
 &\lesssim C_G\frac{\mathcal R_DT_cs_{\rm ref}(T_{\rm osc})}{\mathcal X_D}
 \simeq1.9\times10^{-7}C_G.
 \end{aligned}
 \label{eq:confined-curvature-bound}
\end{equation}
Primes here denote derivatives with respect to the dimensionless gauge angle. The last estimate uses the energy bound, with a density-weighted response coefficient if several species survive. In thermal equilibrium it is also the leading low-density curvature of the free energy, up to corrections of order $T_D/m_G$. For a response coefficient of order unity the thermal or finite-density correction is small enough for the constant-mass calculation. The coefficient must come from the angle-dependent glueball spectrum for a quantitative prediction. This argument uses the actual confined population and its energy bound; it does not assign the extrapolated deconfined temperature to glueballs. The coherent axion potential energy is followed separately and is excluded from $\rho_G$.

The gauge vacuum contributes a potential even when no gas of real particles is present. A particle gas adds an energy density that depends on its number density. Define the yield $Y_\chi=n_\chi/s_{\rm ref}$, the particle number density divided by the visible entropy density. For nonrelativistic axions,
\begin{equation}
 \frac{\rho_{\chi,\rm part}}{s_{\rm ref}}\simeq m_{\chi,0}Y_\chi,
 \qquad Y_{\chi,\rm all\ DM}\simeq1.46\times10^{19}
 \label{eq:particle-yield}
\end{equation}
at the chosen mass. An ordinary relativistic thermal relic has a much smaller yield. Axions produced with significant momentum must become nonrelativistic before their energy can be counted as cold matter.

The inflationary example of Section~\ref{sec:inflation-embedding} gives $\beta_{\rm iso}\simeq0.029$ for the coherent dark matter. Its hidden Higgs vacuum stays stable under the specified inflaton and curvature couplings, and its reheating fractions satisfy Eq.~\eqref{eq:s5-branching}. Other histories must recalculate the phase scale during inflation and the conversion of angle fluctuations into density fluctuations in Eq.~\eqref{eq:s5-iso-power}.

The axion is too light to decay into the heavy portal particles or hidden glueballs. Its decay into visible particles must be calculated with all fields and interactions required for gauge consistency. To survive until today it must satisfy
\begin{equation}
 \Gamma_\chi t_0\ll1.
 \label{eq:survival}
\end{equation}
A photon interaction $-g_{\chi\gamma\gamma}\chi F_{\rm em}\widetilde F_{\rm em}/4$ would give $\Gamma_{\gamma\gamma}=g_{\chi\gamma\gamma}^2m_{\chi,0}^3/(64\pi)$. A coupling of order $1/F_\chi$ gives a negligible width at the chosen mass. The actual coefficient must be calculated in the full theory. The axi-Higgs Yukawa interactions displayed here involve hidden fermions.

For the assumed cosine potential, the magnitude of the quartic contact amplitude is $|\mathcal M|=m_{\chi,0}^2/F_\chi^2$. Define the total cross section by integrating over the full final-state solid angle and including the factor $1/2!$ for two identical outgoing scalars. With the squared centre-of-mass energy $\mathsf s\simeq4m_{\chi,0}^2$, this gives
\begin{equation}
 \sigma_{\chi\chi}=\frac1{2!}\frac1{64\pi^2\mathsf s}
 \int\dd\Omega\,|\mathcal M|^2
 \simeq\frac{m_{\chi,0}^2}{128\pi F_\chi^4}.
 \label{eq:self}
\end{equation}
The large $F_\chi$ makes direct axion scattering extremely weak. The cited Lyman-$\alpha$ analysis obtains $m_{\chi,0}>2\times10^{-20}\,\eV$ at 95\% credibility for an all-dark-matter canonical ultralight axion under its assumed thermal and cosmological model \cite{Lyman}. Our $3\times10^{-20}\,\eV$ example is above that particular limit by a factor $1.5$. This comparison does not replace a fit including our primordial isocurvature contribution. At a speed of $10\,\mathrm{km\,s^{-1}}$, the reduced de Broglie length $\hbar/(m_{\chi,0}v)$ is about $6.4\,\mathrm{pc}$, or $40\,\mathrm{pc}$ when defined with $h/(m_{\chi,0}v)$. The benchmark therefore does not demonstrate kiloparsec wave-supported cores. Its purpose is to illustrate a gauge origin and conditional density for a mass near the quoted structure bound.

A self-interacting dark component would require an additional populated light species and a mediator with suitable mass and couplings. The heavy hidden fermions and portal vector in the present examples do not provide that component. We return to this extension in the outlook.

\section{An anomaly-free generalized \texorpdfstring{$B-L$}{B-L} model}
\label{sec:model2}
We now give a second model in which the fermion charges cancel all gauge anomalies. It has a separate hidden particle spectrum. The physical axion comes from two charged Higgs phases. In both models the hidden non-Abelian gauge theory generates its potential. Related models with a gauged symmetry and a surviving anomalous global phase are discussed in Ref.~\cite{GaugedPQ}. We use the same symbols $X_\mu$ and $g_X$ for the Abelian portal field and coupling in this separate model; their numerical values need not equal those in the first model.

The gauge group and portal generator are
\begin{equation}
 G_2=G_{\rm SM}\times SU(2)_D\times U(1)_X,
 \qquad X=(B-L)_{\rm vis}+D_{\rm hid}.
 \label{eq:model2-group}
\end{equation}
Visible particles have $D_{\rm hid}=0$, so quarks have $X=1/3$, leptons including the three right-handed neutrinos have $X=-1$, and the visible Higgs has $X=0$. Hidden particles have zero visible baryon and lepton numbers, so their $X$ charge is $D_{\rm hid}$. A single Abelian vector couples to these charges with coupling $g_X$. The $X$ symmetry commutes with $SU(2)_D$, so both components of a hidden doublet have the same $X$ charge. Its visible and hidden current interactions are
\begin{equation}
 \mathcal L_{\rm currents,2}
 =g_XX_\mu\left(J_{B-L,\rm vis}^\mu+J_{D,\rm hid}^\mu\right).
 \label{eq:model2-currents}
\end{equation}
These interactions follow from the covariant kinetic terms. The model also allows hypercharge kinetic mixing. We specify its value at a renormalization scale; visible loops generate its running as in the first model.

Introduce two complex scalars that are singlets under $SU(2)_D$ and two hidden Dirac doublets. Table~\ref{tab:model2-fields} gives their representations and charges, separating the visible and hidden contributions to $X$.

\begin{table}[htbp]
\centering
\small
\setlength{\tabcolsep}{4pt}
\renewcommand{\arraystretch}{1.18}
\begin{tabular}{@{}lcccccccc@{}}
\toprule
Field & Spin & $SU(3)_c$ & $SU(2)_L$ & $Y$ & $SU(2)_D$
 & $(B-L)_{\rm vis}$ & $D_{\rm hid}$ & $X$\\
\midrule
$Q_{1L}$ & $\tfrac12$ & $\mathbf1$ & $\mathbf1$ & $0$ & $\mathbf2$ & $0$ & $+1$ & $+1$\\
$Q_{1R}$ & $\tfrac12$ & $\mathbf1$ & $\mathbf1$ & $0$ & $\mathbf2$ & $0$ & $-1$ & $-1$\\
$Q_{2L}$ & $\tfrac12$ & $\mathbf1$ & $\mathbf1$ & $0$ & $\mathbf2$ & $0$ & $-1$ & $-1$\\
$Q_{2R}$ & $\tfrac12$ & $\mathbf1$ & $\mathbf1$ & $0$ & $\mathbf2$ & $0$ & $+1$ & $+1$\\
$\Phi_1$ & $0$ & $\mathbf1$ & $\mathbf1$ & $0$ & $\mathbf1$ & $0$ & $+2$ & $+2$\\
$\Phi_2$ & $0$ & $\mathbf1$ & $\mathbf1$ & $0$ & $\mathbf1$ & $0$ & $-2$ & $-2$\\
\bottomrule
\end{tabular}
\caption{Hidden fields of the second model before symmetry breaking. The shared charge is $X=(B-L)_{\rm vis}+D_{\rm hid}$. Every hidden field has $(B-L)_{\rm vis}=0$, so its $X$ charge is $D_{\rm hid}$. The two Dirac doublets have opposite left- and right-handed portal-charge assignments.}
\label{tab:model2-fields}
\end{table}

We impose the global phase symmetry
\begin{equation}
 \Phi_i\mapsto e^{i\alpha}\Phi_i,
 \qquad Q_{iL}\mapsto Q_{iL},\qquad Q_{iR}\mapsto e^{-i\alpha}Q_{iR},
 \qquad i=1,2.
 \label{eq:model2-global}
\end{equation}
Visible fields are unchanged by this symmetry. It allows the Yukawa interactions below and forbids perturbative terms that would give the remaining phase a mass. For example, $\Phi_1\Phi_2$ has global charge two and cannot appear alone in the symmetric potential. The symmetry also forbids off-diagonal fermion mass terms and Yukawa interactions with the conjugate scalars. Its hidden non-Abelian anomaly generates the intended axion potential. Its protection at higher energies remains an independent assumption. In particular, the gauge-invariant product $\Phi_1\Phi_2$ plays the same role as $S^\dagger\phi$ in Section~\ref{sec:phase-protection}. For a series $(\Phi_1\Phi_2)^n$ with order-one coefficients and cutoff $\mpl$, the equal-vev benchmark has $v_1v_2/(2\mpl^2)=0.00166$. The one-percent curvature criterion then requires the first allowed power to satisfy $n\geq37$, corresponding to dimension at least $74$. A discrete gauge protection would need its own consistent anomaly conditions; gauge-anomaly cancellation of the displayed continuous group does not enforce this suppression.

Before choosing the vacuum, write the scalar potential as
\begin{equation}
 \begin{aligned}
 V_2(\Phi_1,\Phi_2)={}&
 -\mu_1^2\Phi_1^\dagger\Phi_1-\mu_2^2\Phi_2^\dagger\Phi_2
 +\lambda_1(\Phi_1^\dagger\Phi_1)^2\\
 &+\lambda_2(\Phi_2^\dagger\Phi_2)^2
 +\lambda_{12}(\Phi_1^\dagger\Phi_1)(\Phi_2^\dagger\Phi_2).
 \end{aligned}
 \label{eq:model2-potential}
\end{equation}
This potential respects both the $X$ gauge symmetry and the global symmetry in Eq.~\eqref{eq:model2-global}. It is independent of the two scalar phases in perturbation theory.

The hidden action, including the shared Abelian vector, is
\begin{equation}
 \begin{aligned}
 \mathcal L_{D,2}={}&-\frac14W^a_{\mu\nu}W^{a\mu\nu}
 -\frac14F_{X\mu\nu}F_X^{\mu\nu}
 +\sum_{i=1}^2|D_\mu\Phi_i|^2
 +\sum_{i=1}^2\bar Q_i i\gamma^\mu D_\mu Q_i\\
 &-\left(y_1\Phi_1\bar Q_{1L}Q_{1R}
 +y_2\Phi_2\bar Q_{2L}Q_{2R}+\mathrm{h.c.}\right)-V_2,
 \end{aligned}
 \label{eq:model2-lagrangian}
\end{equation}
where $F_X=\dd X$, and the hidden non-Abelian field strength uses the coupling $g_{D,2}$. With $q_1=+2$, $q_2=-2$ and the fermion charges in Table~\ref{tab:model2-fields}, the derivatives are
\begin{equation}
 \begin{aligned}
 D_\mu\Phi_i&=(\partial_\mu-ig_Xq_iX_\mu)\Phi_i,\\
 D_\mu Q_i&=\left[\partial_\mu-ig_{D,2}T^aW_\mu^a
 -ig_XX_\mu(x_{iL}P_L+x_{iR}P_R)\right]Q_i.
 \end{aligned}
 \label{eq:model2-derivatives}
\end{equation}
Each Yukawa charge sum is $q_i-x_{iL}+x_{iR}=0$. The coupling $g_{D,2}$ belongs to this second realization.

The gauge-anomaly traces cancel between the two doublets. In the conventions used for the first model,
\begin{equation}
 \begin{aligned}
 C^{(2)}_{XDD}&=\frac12\left[(1-(-1))+(-1-1)\right]=0,\\
 C^{(2)}_{XXX}&=2\left[1^3-(-1)^3+(-1)^3-1^3\right]=0,\\
 C^{(2)}_{X\mathrm{grav}}&=2\left[1-(-1)+(-1)-1\right]=0.
 \end{aligned}
 \label{eq:model2-anomalies}
\end{equation}
The label $XDD$ denotes one $X$ and two $SU(2)_D$ vertices; $XXX$ denotes three $X$ vertices; and $X\mathrm{grav}$ denotes one $X$ and two gravitational vertices. The four left-handed Weyl doublets also satisfy the global $SU(2)$ condition. The visible gauge anomalies cancel when the three right-handed neutrinos are included. The gauge theory is therefore anomaly free. The global phase symmetry still has the hidden anomaly that generates the small angular potential.

For a vacuum with $|\langle\Phi_i\rangle|=v_i/\sqrt2$ and both $v_i>0$, minimizing Eq.~\eqref{eq:model2-potential} gives
\begin{equation}
 \mu_1^2=\lambda_1v_1^2+\frac{\lambda_{12}}2v_2^2,
 \qquad\mu_2^2=\lambda_2v_2^2+\frac{\lambda_{12}}2v_1^2.
 \label{eq:model2-stationarity}
\end{equation}
The two squared vacuum values are
\begin{equation}
 \begin{gathered}
 \Delta_2=\lambda_1\lambda_2-\frac{\lambda_{12}^2}{4},\\
 v_1^2=\frac{\lambda_2\mu_1^2-\lambda_{12}\mu_2^2/2}{\Delta_2},
 \qquad
 v_2^2=\frac{\lambda_1\mu_2^2-\lambda_{12}\mu_1^2/2}{\Delta_2}.
 \end{gathered}
 \label{eq:model2-vevs}
\end{equation}
We choose positive squared vacuum values, $\lambda_1,\lambda_2>0$ and $\Delta_2>0$. The radial squared-mass matrix is
\begin{equation}
 \mathcal M_{h,2}^2=
 \begin{pmatrix}
 2\lambda_1v_1^2&\lambda_{12}v_1v_2\\
 \lambda_{12}v_1v_2&2\lambda_2v_2^2
 \end{pmatrix}.
 \label{eq:model2-radial-masses}
\end{equation}
Its trace and determinant $4v_1^2v_2^2\Delta_2$ are positive, so both radial squared masses are positive. We add the much smaller angular potential after determining this leading vacuum.

We can now separate the two Higgs phases:
\begin{equation}
 \Phi_i=\frac{v_i+h_i}{\sqrt2}e^{i\eta_i/v_i},
 \qquad m_{Q_i}=\frac{y_iv_i}{\sqrt2}.
 \label{eq:model2-polar}
\end{equation}
Their kinetic terms at fixed magnitudes are
\begin{equation}
 \mathcal L_{\rm phase,2}
 =\frac12(\partial\eta_1-2g_Xv_1X)^2
 +\frac12(\partial\eta_2+2g_Xv_2X)^2.
 \label{eq:model2-phase-action}
\end{equation}
The gauge transformation shifts the phases along $(v_1,-v_2)$. With $v=\sqrt{v_1^2+v_2^2}$, define the combinations along this shift and perpendicular to it:
\begin{equation}
 G_X=\frac{v_1\eta_1-v_2\eta_2}{v},\qquad
 \chi_2=\frac{v_2\eta_1+v_1\eta_2}{v},\qquad m_X=2g_Xv.
 \label{eq:model2-physical-phase}
\end{equation}
These fields give $\mathcal L_{\rm phase,2}=\tfrac12(\partial G_X-m_XX)^2+\tfrac12(\partial\chi_2)^2$. The vector absorbs $G_X$ as its longitudinal polarization, leaving $\chi_2$ as a physical scalar with the standard kinetic term. The gauge-invariant scalar product and its angle are
\begin{equation}
 \Phi_1\Phi_2=\frac{(v_1+h_1)(v_2+h_2)}2 e^{i\Theta_2},
 \qquad\Theta_2=\frac{\eta_1}{v_1}+\frac{\eta_2}{v_2}
 =\frac{\chi_2}{F_2},\qquad F_2=\frac{v_1v_2}{\sqrt{v_1^2+v_2^2}}.
 \label{eq:model2-angle}
\end{equation}
The gauge shifts cancel in the sum of the two phases. The global symmetry in Eq.~\eqref{eq:model2-global} changes this sum by $2\alpha$, so the surviving physical angle carries that global symmetry.

Both hidden fermion masses contribute a phase to the gauge theta angle. Chiral field redefinitions that make these masses real give
\begin{equation}
 \mathcal L_{\rm top,2}
 =\left(\theta_{D,2}+\frac{\eta_1}{v_1}+\frac{\eta_2}{v_2}\right)Q_D
 =\left(\theta_{D,2}+\frac{\chi_2}{F_2}\right)Q_D,
 \qquad Q_D=\frac{g_{D,2}^2}{32\pi^2}W^a_{\mu\nu}\widetilde W^{a\mu\nu}.
 \label{eq:model2-topological}
\end{equation}
Each Dirac doublet contributes one mass phase. Their sum is gauge invariant and has the global anomaly needed to generate the potential. The angular energy depends on $\arg(\Phi_1\Phi_2)$ and the two Higgs magnitudes. Its second angular derivative at the vacuum gives
\begin{equation}
 m_{\chi_2}^2(T_D)=\frac{\mathcal X_{D,2}(T_D)}{F_2^2}.
 \label{eq:model2-mass}
\end{equation}
The full angle dependence and its temperature dependence require a strong-coupling calculation, as in the first model.

Above both fermion masses, the one-loop running coefficient is $b_{(2)}=22/3-2(2/3)=6$. Below both masses it is $b_0=22/3$. Removing each fermion at its mass and joining the running solutions gives
\begin{equation}
 \Lambda_0^{22/3}=m_{Q_1}^{2/3}m_{Q_2}^{2/3}\Lambda_{(2)}^6,
 \label{eq:model2-two-thresholds}
\end{equation}
Here $\Lambda_{(2)}$ is the scale defined with two active Dirac doublets. This equation gives how the low-energy scale depends on the Higgs magnitudes. Holding the high-energy coupling and pure-gauge coefficient fixed gives $\mathcal X_{D,2}\propto[(v_1+h_1)(v_2+h_2)]^{4/11}$ when the fermions are heavy compared with confinement. For $v_1=v_2=v_0$ and equal fermion masses $m_Q$, it reduces to
\begin{equation}
 \Lambda_0=m_Q\exp\left[-\frac1{22/3}
 \left(\frac{2\pi}{\alpha_{D,2}(v_0)}+6\ln\frac{m_Q}{v_0}\right)\right].
 \label{eq:model2-running}
\end{equation}
The relation $\mathcal X_{D,2}(0)=c_0\Lambda_0^4$ again requires $c_0$ to be assumed or obtained from the strong gauge theory.

To compare the two models at the same phase scale, choose
\begin{equation}
 \begin{gathered}
 v_1=v_2=v_0=1.404\times10^{17}\,\GeV,\qquad g_X=0.3,\\
 \lambda_1=\lambda_2=0.01,\qquad\lambda_{12}=0,
 \qquad y_1=y_2=0.2.
 \end{gathered}
 \label{eq:model2-benchmark}
\end{equation}
These values give
\begin{equation}
 \begin{gathered}
 F_2=9.929\times10^{16}\,\GeV,\qquad m_X=1.191\times10^{17}\,\GeV,\\
 m_{Q_1}=m_{Q_2}=m_{h_1}=m_{h_2}=1.986\times10^{16}\,\GeV.
 \end{gathered}
 \label{eq:model2-benchmark-scales}
\end{equation}
In the one-loop $c_0=1$ illustration, $\Lambda_0=1.726\,\mathrm{keV}$ and $\alpha_{D,2}(v_0)\simeq0.0164$ give $m_{\chi_2}=3\times10^{-20}\,\eV$. Using the same lattice normalization and the gauge--fermion two-loop running with $n_f=2$ above the common fermion threshold gives $\alpha_{D,2}^{\overline{\rm MS}}(v_0)\simeq0.0159$ for that target. For the same cosine potential and cold history, the uniform field has the same mass and phase scale as in the first example. The initial angle $\theta_i\simeq0.325$ therefore again gives $\Omega_{\rm coh}h_0^2\simeq0.12$.

The vacuum test in Appendix~\ref{app:s5-scalar-minima} also applies here, with $\phi,S$ replaced by $\Phi_1,\Phi_2$ and the quartic matrix of this model. For the equal-vev example, the same small inflaton and curvature couplings preserve both nonzero vacuum values, and both radial masses remain much larger than the Hubble rate. Appendix~\ref{app:s5-temperature} gives the squared-vev shifts, the change of $F_2$ and the thermal bound. The physical phase therefore survives the stated history with the same isocurvature power to the quoted accuracy. The small hidden reheating fraction and protection of the global phase remain assumptions.

The different vector and fermion masses change the particle-production rates. At $T_R=10^{10}\,\GeV$, the current-interaction estimate gives $\Gamma_X/H\sim2.9\times10^{-23}C$; producing the heavy charged particles is further suppressed. The two sectors can therefore keep different temperatures in this example. Their initial entropy ratio, survival of the Higgs vacuum and protection of the global phase must still be specified. Visible decay amplitudes must be calculated using this model's spectrum before applying Eq.~\eqref{eq:survival}. This gives a simple gauge-anomaly-free model of the generalized $B-L$ portal with an ultralight Higgs phase.

\section{Conclusions and outlook}
\label{sec:conclusions}
The charged hidden Higgs phase and the St\"uckelberg field leave one physical scalar after the Abelian vector acquires its longitudinal polarization. We derive its phase scale and the hidden gauge angle. A second construction uses two charged Higgs singlets and cancels gauge anomalies within its displayed fermion spectrum. Both constructions admit a large physical phase scale with small internal representations.

The gauge charge assignments also permit low-dimensional scalar interactions that would give this phase a large mass. Our ultralight examples therefore depend on an imposed perturbative symmetry and its preservation at higher energies. The explicit curvature bounds show how strong that protection must be. A complete discrete gauge mechanism or another high-energy explanation remains to be constructed. The gauge pathway specifies the field and anomaly-induced potential within this conditional effective theory.

The susceptibility relates the physical mass to the phase scale. Pure-$SU(2)$ lattice ratios connect it to the string tension and transition temperature, and a representative conversion to the $\overline{\rm MS}$ scale improves on an arbitrary one-loop normalization. The required high-energy coupling depends appreciably on the running approximation. The mass is selected through that coupling, and the initial angle is selected through the density relation. Our example uses $F_\chi\simeq9.93\times10^{16}\,\GeV$, $m_{\chi,0}=3\times10^{-20}\,\eV$ and $\theta_i\simeq0.325$. The analytic radiation-era solution and the assumed cosine evolution then give the reference coherent density. The remaining strong-dynamics and threshold uncertainties prevent a precision prediction of these selected inputs.

The cold confined population obeys separate number and energy equations. Number-changing reactions can keep it warmer than a redshifted radiation bath, while its energy in an expanding comoving region still decreases for nonnegative pressure. Conserved hidden entropy through equilibrium confinement gives a late energy bound below $9\times10^{-6}$ of the reference dark-matter density in the example. Its contribution to the angular curvature is also small for an ordinary glueball mass response. These results state the entropy, isolation and spectrum assumptions needed for the constant-mass treatment.

The illustrative inflaton background passes the stated scalar-vacuum and isocurvature checks. Its CMB tilt comparison depends on the reheating pressure and on fitting isocurvature jointly with the scalar spectrum. Portal-induced quantum corrections require protection or adjustment of the renormalized inflaton potential. Moreover, keeping the hidden sector cold demands an inflaton coupling to hidden gauge bosons about a million times weaker than its coupling to visible hypercharge. The displayed symmetries do not produce this hierarchy. Microscopic reheating, hidden thermalization and the high-energy protection of both potentials remain necessary additions.

Gravothermal black-hole seeding requires a populated collisional component, beyond the almost collisionless axion in these examples. If a lighter species $d$ and its mediator are added, their production must determine the matter fraction and halo scattering rate,
\begin{equation}
 \begin{gathered}
 f_d=\frac{\Omega_d}{\Omega_{\rm DM}},\qquad
 n_d=\frac{\rho_d}{m_d},\qquad
 \Gamma_{\rm el}=n_d\langle\sigma_{dd}v_{\rm rel}\rangle,\\
 \Omega_{\rm DM}=\Omega_{\rm coh}+\Omega_{\chi,\rm part}+\Omega_d+\Omega_{\rm other}.
 \end{gathered}
 \label{eq:interacting-fraction}
\end{equation}
Here $\sigma_{dd}$ is the total elastic cross section. Heat transport also requires its angular and velocity dependence \cite{HaloTransport}. The subsequent collapse depends on the interacting fraction, baryons and the change from a long to a short mean free path; recent calculations find substantially slower late collapse in the optically thick regime \cite{CollapseSeeds,EarlySeeds,BaryonicSeeds,CollapseCalibration}. Perturbation growth, halo evolution, relativistic collapse and later accretion must all be calculated before connecting such an extension to little red dots \cite{LRDSeeds,Accretion}. The present results provide the conditional axion background for that further analysis.

\FloatBarrier
\section*{Acknowledgements}
This work is supported by INFN, Iniziativa Specifica QGsky.

\appendix
\numberwithin{equation}{section}
\section{Hidden vacuum energy and its Higgs dependence}
\label{app:vacuum-energy}
At fixed $\rho_\phi=\phi^\dagger\phi$, the hidden gauge vacuum depends on the Higgs magnitude through $m_F=y_F\sqrt{\rho_\phi}$. We derive Eq.~\eqref{eq:angle-independent-vacuum-energy} with the high-energy gauge coupling held fixed and $m_F\gg\Lambda_D$. After the chiral redefinition in Section~\ref{sec:gauge-mass}, the renormalized vacuum energy is defined by
\begin{equation}
 \begin{aligned}
 Z_D(\rho_\phi,\vartheta)
 &=\int\mathcal D W\,\mathcal D\bar F\,\mathcal D F\,
 \exp\!\left[-S_{D,E}(\rho_\phi)+i\vartheta\nu_D\right],\\
 \mathcal E_D^{\rm full}(\rho_\phi,\vartheta)
 &=-\lim_{\mathcal V_4\to\infty}
 \frac{1}{\mathcal V_4}\ln Z_D(\rho_\phi,\vartheta)
 \bigg|_{\rm ren}.
 \end{aligned}
 \label{eq:app-vacuum-functional}
\end{equation}
Here $S_{D,E}$ is the Euclidean gauge and fermion action, $\nu_D$ is the integer topological charge, and $\mathcal V_4$ is the Euclidean four-volume. The superscript retains the perturbative angle-independent contributions explicitly.

The one-loop coefficient with $j$ active Dirac doublets is \cite{GaugeBetaFunction}
\begin{equation}
 b_j=\frac{11}{3}C_A-\frac{4}{3}T_Fj
 =\frac{22}{3}-\frac{2j}{3},
 \qquad b_1=\frac{20}{3},\quad b_0=\frac{22}{3}.
 \label{eq:app-beta-content}
\end{equation}
For $SU(2)$, the adjoint Casimir is $C_A=2$, while $T^a=\sigma^a/2$ gives $\tr(T^aT^b)=T_F\delta^{ab}$ with $T_F=1/2$. Thus the gauge fields contribute $22/3$ and the single Dirac doublet subtracts $2/3$. The scalars and $\Psi$ are hidden gauge singlets. With $\alpha_j=g_j^2/(4\pi)$, the beta function and its integrated form are
\begin{equation}
 \mu\frac{\dd g_j}{\dd\mu}=-\frac{b_j}{16\pi^2}g_j^3,
 \qquad
 \alpha_j^{-1}(\mu)=\frac{b_j}{2\pi}
 \ln\frac{\mu}{\Lambda_j}.
 \label{eq:app-running-scales}
\end{equation}
The scales $\Lambda_j$ are integration constants. The one-flavour expression applies above $m_F$, and the pure-gauge expression applies for $\Lambda_D\ll\mu<m_F$, with $\Lambda_0=\Lambda_D(\rho_\phi)$. Continuity at $\mu=m_F$ gives \cite{HeavyDecoupling,HeavyMassScaling}
\begin{equation}
 b_0\ln\frac{m_F}{\Lambda_D}
 =b_1\ln\frac{m_F}{\Lambda_1},
 \qquad
 \Lambda_D^{22/3}=m_F^{2/3}\Lambda_1^{20/3}.
 \label{eq:app-threshold-matching}
\end{equation}
Consequently,
\begin{equation}
 \Lambda_D(\rho_\phi)
 =\Lambda_1\left(\frac{y_F\sqrt{\rho_\phi}}{\Lambda_1}\right)^{1/11},
 \qquad
 \frac{\Lambda_D^4(\rho_\phi)}{\Lambda_D^4(\rho_*)}
 =\left(\frac{\rho_\phi}{\rho_*}\right)^{2/11},
 \quad \rho_*=\frac{v_\phi^2}{2}.
 \label{eq:app-radial-scale}
\end{equation}
The exponent follows from $(b_0-b_1)/b_0=1/11$. Running of $y_F$ and higher-order threshold corrections modify this leading result.

Below the doublet mass, the only dynamical scale of the pure gauge theory is $\Lambda_D$. Its confinement contribution to an energy density therefore has the form $-c_E\Lambda_D^4$, up to terms suppressed by powers of $\Lambda_D/m_F$. Including the perturbative contribution gives
\begin{equation}
 \mathcal E_D^{\rm full}(\rho_\phi,0)
 \simeq\mathcal E_{\rm ref}
 +V_D^{\rm pert}(\rho_\phi;\mu)
 -c_E\Lambda_D^4(\rho_\phi).
 \label{eq:app-full-radial-energy}
\end{equation}
The signed coefficient $c_E$ requires nonperturbative input in a specified subtraction prescription \cite{VacuumEnergySubtraction}; $\mathcal E_{\rm ref}$ fixes the additive normalization. The one-loop effective potential, its doublet multiplicity and the associated mass responses are given in Appendix~\ref{app:s5-loops}.
Assigning $V_D^{\rm pert}$ to the radial effective potential leaves $\mathcal E_{D,0}=\mathcal E_{\rm ref}-c_E\Lambda_D^4$. Subtraction at the broken vacuum gives
\begin{equation}
 \mathcal E_{D,0}(\rho_\phi)-\mathcal E_{D,0}(\rho_*)
 =-c_E\Lambda_D^4(\rho_*)
 \left[\left(\frac{\rho_\phi}{\rho_*}\right)^{2/11}-1\right].
 \label{eq:app-subtracted-radial-energy}
\end{equation}
Including all angle-independent terms in $V_{\rm rad}$ yields Eq.~\eqref{eq:covariant-hidden-potential}. Its derivatives determine the radial vacuum and mass; Eq.~\eqref{eq:radial} gives their leading polynomial description.

The angular contribution is defined by subtracting the energy at the same Higgs magnitude and zero angle. In the leading-harmonic approximation,
\begin{equation}
 \begin{aligned}
 \mathcal U_D(\rho_\phi,\vartheta)
 &\simeq c_\chi\Lambda_D^4(\rho_\phi)
 [1-\cos\vartheta],\qquad c_\chi>0,\\
 \mathcal X_D(0)&=c_\chi\Lambda_D^4(\rho_*),
 \qquad m_{\chi,0}^2=\frac{c_\chi\Lambda_D^4(\rho_*)}{F_\chi^2}.
 \end{aligned}
 \label{eq:app-angular-energy}
\end{equation}
The coefficients $c_\chi$ and $c_E$ require separate nonperturbative inputs. Higher angular harmonics also depend on the confining dynamics \cite{SU2Topology}. These expressions assume a heavy doublet and a slowly varying Higgs field; the region $m_F\sim\Lambda_D$ requires the theory with the fermion retained.

\input{appendices/Section5_Derivations}

\FloatBarrier
\begingroup
\small
\setlength{\bibsep}{3pt}
\interlinepenalty=10000

\endgroup
\end{document}

%% file: appendices/Section5_Derivations.tex
\section{Partial decoupling of the scalar S}
\label{app:s5-scalar-decoupling}
The scalar realization begins with an $SU(2)_D$ singlet $S$ of portal charge two \cite{ScalarDecoupling}. Its action contains
\begin{equation}
 \mathcal L_S=|D_\mu S|^2
 -\lambda_S\left(|S|^2-\frac{w^2}{2}\right)^2,
 \qquad D_\mu S=(\partial_\mu-ie_XX_\mu)S.
 \label{eq:s5-S-action}
\end{equation}
Writing $S=(w+s)e^{ib/w}/\sqrt2$ gives
\begin{equation}
 D_\mu S=\frac{e^{ib/w}}{\sqrt2}
 \left[\partial_\mu s+i\left(1+\frac{s}{w}\right)
 (\partial_\mu b-e_XwX_\mu)\right].
 \label{eq:s5-S-derivative}
\end{equation}
The real and imaginary terms have zero cross term in the modulus squared, so
\begin{equation}
 |D_\mu S|^2=\frac12(\partial s)^2
 +\frac12\left(1+\frac{s}{w}\right)^2
 (\partial b-M_1X)^2,\qquad M_1=e_Xw.
 \label{eq:s5-S-kinetic}
\end{equation}
Expanding the potential gives
\begin{equation}
 V_S=\lambda_Sw^2s^2+\lambda_Sws^3+\frac{\lambda_S}{4}s^4,
 \qquad m_s^2=2\lambda_Sw^2.
 \label{eq:s5-radial-mass}
\end{equation}
The potential is independent of $b$. A large $w$ at fixed $\lambda_S$ raises the radial mass while preserving the angular direction. This is the separation between a heavy radial excitation and a Goldstone phase familiar from the Peccei--Quinn mechanism \cite{PQMechanism}. Gauge interactions subsequently determine which phase combination is physical.

The integration of the radial mode can be performed explicitly. Define $K_b=(\partial b-M_1X)^2$. Keeping the terms needed at leading order gives
\begin{equation}
 \mathcal L_S=\frac12K_b+\frac12(\partial s)^2
 -\frac12m_s^2s^2+\frac{s}{w}K_b+\cdots,
 \qquad (\Box+m_s^2)s=\frac{K_b}{w}+\cdots.
 \label{eq:s5-radial-eom}
\end{equation}
For momenta much smaller than $m_s$, the solution is $s\simeq K_b/(wm_s^2)$. Substitution then gives
\begin{equation}
 \mathcal L_{\rm eff}=\frac12K_b
 +\frac{K_b^2}{2w^2m_s^2}+\cdots.
 \label{eq:s5-radial-elimination}
\end{equation}
The first term is the St\"uckelberg kinetic term, while the next term displays the suppression of the omitted radial dynamics. This partial decoupling integrates over fluctuations of the magnitude about $w$. The expectation value, the phase and the parameters inherited from $w$ remain in the low-energy action.

Under $X_\mu\to X_\mu+\partial_\mu\omega$, the phases transform as
\begin{equation}
 b\to b+e_Xw\omega,\qquad
 \eta\to\eta+e_Xv_\phi\omega,
 \qquad \phi=\frac{v_\phi+h}{\sqrt2}e^{i\eta/v_\phi}.
 \label{eq:s5-phase-transformations}
\end{equation}
With both magnitudes fixed, the phase kinetic terms are diagonalized by
\begin{equation}
 G=\frac{v_\phi\eta+wb}{\sqrt{v_\phi^2+w^2}},\qquad
 \chi=\frac{v_\phi b-w\eta}{\sqrt{v_\phi^2+w^2}},\qquad
 \mathcal L_{\rm phase}=\frac12(\partial G-M_XX)^2
 +\frac12(\partial\chi)^2.
 \label{eq:s5-phase-rotation}
\end{equation}
Thus $\delta G=M_X\omega$ and $\delta\chi=0$. The vector absorbs $G$. The physical field is the relative phase,
\begin{equation}
 \Theta=\frac{\eta}{v_\phi}-\frac{b}{w}
 =-\frac{\chi}{F_\chi},\qquad
 F_\chi^{-2}=v_\phi^{-2}+w^{-2}.
 \label{eq:s5-relative-phase}
\end{equation}
An exact relative-phase symmetry makes $\chi$ massless in perturbation theory; the hidden anomaly supplies its eventual small potential. With a single gauged scalar the phase instead supplies the vector's longitudinal polarization. In the hierarchy $w\gg v_\phi$, $G\simeq b$, $\chi\simeq-\eta$ and $F_\chi\simeq v_\phi$. Taking $w$ large also raises $M_X$ at fixed $e_X$; at energies below both heavy masses the vector can be integrated out as well.

For the numerical example, $m_s\simeq8.33\times10^{17}\,\GeV$, $M_X\simeq5.04\times10^{17}\,\GeV$ and $H_*\simeq7.96\times10^{11}\,\GeV$. Inflation is therefore far below both heavy scales. Retaining the radial variables temporarily to find their displaced minima gives a consistent tree-level matching calculation. The resulting low-energy evolution uses the surviving relative phase.

\section{Inflationary background and expansion}
\label{app:s5-inflation}
\label{sec:inflation-background}
We take a canonical neutral inflaton $I$ with potential \cite{PolynomialInflation}
\begin{equation}
 U(I)=U_0\frac{I^2}{I^2+\mu_I^2}.
 \label{eq:s5-potential}
\end{equation}
The derivatives entering the background and perturbation equations follow directly:
\begin{equation}
 \begin{aligned}
 U'&=\frac{2U_0I\mu_I^2}{(I^2+\mu_I^2)^2},\quad
 U''=\frac{2U_0\mu_I^2(\mu_I^2-3I^2)}{(I^2+\mu_I^2)^3},\\
 \frac{U'}{U}&=\frac{2\mu_I^2}{I(I^2+\mu_I^2)},\qquad
 \frac{U''}{U}=\frac{2\mu_I^2(\mu_I^2-3I^2)}{I^2(I^2+\mu_I^2)^2}.
 \end{aligned}
 \label{eq:s5-potential-derivatives}
\end{equation}
The energy density and pressure of a homogeneous canonical scalar are $\rho_I=\dot I^2/2+U$ and $p_I=\dot I^2/2-U$. Energy conservation, $\dot\rho_I+3H(\rho_I+p_I)=0$, gives the inflaton equation. Together with the Friedmann equations,
\begin{equation}
 \ddot I+3H\dot I+U'=0,\qquad
 3\mpl^2H^2=\frac12\dot I^2+U,\qquad
 \dot H=-\frac{\dot I^2}{2\mpl^2}.
 \label{eq:s5-background}
\end{equation}
The slow-roll approximation neglects $\ddot I$ relative to $3H\dot I$ and the kinetic energy relative to $U$. It gives $3H\dot I\simeq-U'$ and $3\mpl^2H^2\simeq U$. The potential parameters are consequently
\begin{equation}
 \epsilon_V=\frac{2\mpl^2\mu_I^4}{I^2(I^2+\mu_I^2)^2},\qquad
 \eta_V=\frac{2\mpl^2\mu_I^2(\mu_I^2-3I^2)}{I^2(I^2+\mu_I^2)^2}.
 \label{eq:s5-slowroll}
\end{equation}
The number of e-folds, $N=\int H\dd t$, becomes
\begin{equation}
 \begin{aligned}
 N_{\rm sr}&=\frac1{\mpl^2}\int_{I_e}^{I_*}\frac{U}{U'}\dd I
 =\frac1{2\mu_I^2\mpl^2}\int_{I_e}^{I_*}(I^3+\mu_I^2I)\dd I\\
 &=\frac{I_*^4-I_e^4}{8\mu_I^2\mpl^2}
 +\frac{I_*^2-I_e^2}{4\mpl^2}.
 \end{aligned}
 \label{eq:s5-Nsr}
\end{equation}
For $I_*\gg\mu_I$, this gives $I_*^4\simeq8N_{\rm sr}\mu_I^2\mpl^2$. Inserting this relation into $n_s\simeq1-6\epsilon_V+2\eta_V$ gives the leading plateau result $n_s\simeq1-3/(2N_{\rm sr})$. Near the end, we solve the full background equations and use the first crossing of $\epsilon_H=-\dot H/H^2=1$.

The first-order primordial spectra are \cite{Inflation}
\begin{equation}
 \mathcal P_{\mathcal R}=\frac{H_*^2}{8\pi^2\mpl^2\epsilon_{H*}},\qquad
 n_s-1=-2\epsilon_{H*}-\epsilon_{2*},\qquad
 r=16\epsilon_{H*},\quad
 \epsilon_2=\frac{\dd\ln\epsilon_H}{\dd\ln a}.
 \label{eq:s5-spectra}
\end{equation}
The amplitude $\mathcal P_{\mathcal R}=2.10\times10^{-9}$ therefore fixes
\begin{equation}
 \epsilon_{H*}=\frac{H_*^2}{8\pi^2\mpl^2\mathcal P_{\mathcal R}}
 =6.4440\times10^{-7}
 \quad\hbox{for}\quad H_*=7.9594\times10^{11}\,\GeV.
 \label{eq:s5-amplitude-number}
\end{equation}
The corresponding energy scale is $(3\mpl^2H_*^2)^{1/4}=1.8322\times10^{15}\,\GeV$. Both quantities have mass dimension one; the Planck scale in the Friedmann equation makes their numerical values very different.

For numerical integration, define $y=I/\mpl$, $s_I=\mu_I/\mpl$, $\tau=\sqrt{U_0}t/\mpl$, $z=\dd y/\dd\tau$, and $q(y)=y^2/(y^2+s_I^2)$. Then
\begin{equation}
 \frac{\dd y}{\dd\tau}=z,\qquad
 \frac{\dd z}{\dd\tau}=-3\mathfrak h z-q_y,\qquad
 \frac{\dd N}{\dd\tau}=\mathfrak h,\qquad
 \mathfrak h=\sqrt{\frac{z^2/2+q}{3}}.
 \label{eq:s5-dimensionless-background}
\end{equation}
Here $H=\sqrt{U_0}\mathfrak h/\mpl$ and $\epsilon_H=z^2/(2\mathfrak h^2)$. We integrate the homogeneous equations from a point slightly before the reference field value on the slow-roll attractor through the first oscillations, identifying the pivot and first end of inflation separately. The amplitude fixes
\begin{equation}
 U_0=\frac{8\pi^2\mathcal P_{\mathcal R}\mpl^4\epsilon_{H*}}
 {\mathfrak h_*^2}.
 \label{eq:s5-normalization}
\end{equation}
At the pivot, $\epsilon_{2*}=0.0305998$. Thus $n_s=0.969399$ and $r=1.0310\times10^{-5}$. The potential estimate gives $\eta_V=-0.0154802$ and $n_s=0.969036$; their small difference measures the use of potential slow roll versus the evolved first-order Hubble parameters.

To connect this evolution to the observed pivot $k_*=0.05\,\mathrm{Mpc}^{-1}$, write $k_*=a_*H_*$. Choose a matching time $t_m$ during oscillations and the end of reheating $t_R$. The scale factors obey
\begin{equation}
 \frac{k_*}{a_0H_*}
 =\frac{a_*}{a_m}\frac{a_m}{a_R}\frac{a_R}{a_0}
 =e^{-N_{*m}-N_{mR}}\frac{T_0}{T_R}
 \left(\frac{g_{*s,0}}{g_{*s,R}}\right)^{1/3}.
 \label{eq:s5-pivot-matching}
\end{equation}
The last factor follows from conservation of visible entropy, $g_{*s}T^3a^3=\mathrm{constant}$, after reheating. We use $T_0=2.34865\times10^{-13}\,\GeV$, $g_{*s,0}=43/11$, $g_{*s,R}=106.75$ and $T_R=10^{10}\,\GeV$. In energy units $k_*=3.19747\times10^{-40}\,\GeV$. Taking logarithms gives the target expansion
\begin{equation}
 N_{*m}+N_{mR}
 =\ln\frac{H_*}{k_*}+\ln\frac{T_0}{T_R}
 +\frac13\ln\frac{g_{*s,0}}{g_{*s,R}}
 =65.1358.
 \label{eq:s5-target-N}
\end{equation}
This includes expansion after the first end of inflation.

The oscillation-averaged pressure supplies that later expansion. At fixed oscillation energy $E=eU_0< U_0$, the amplitude is $I_{\rm amp}=\mu_I\sqrt{e/(1-e)}$. Set $A^2=e/(1-e)$ and $I=\mu_IA\sin u$. The period averages of the kinetic energy give
\begin{equation}
 \overline w_I(e)=2\frac{J_1(e)}{J_0(e)}-1,\qquad
 J_0=\int_0^{\pi/2}\sqrt{1+A^2\sin^2u}\dd u,\quad
 J_1=\int_0^{\pi/2}\frac{\cos^2u}{\sqrt{1+A^2\sin^2u}}\dd u.
 \label{eq:s5-averaged-pressure}
\end{equation}
Indeed, $\overline w_I=2\langle K\rangle/E-1$, and the numerator and denominator above arise from integrating $K\dd t$ and $\dd t$ over a cycle. For $e\ll1$, expansion of the integrals gives $\overline w_I=-3e/8+\cdots$, approaching the pressureless quadratic limit. Energy conservation then gives
\begin{equation}
 N_{mR}=\frac13\int_{\rho_R}^{\rho_m}
 \frac{\dd\rho_I}{\rho_I[1+\overline w_I(\rho_I)]},\qquad
 \rho_R=\frac{\pi^2g_{*,R}}{30}T_R^4.
 \label{eq:s5-reheating-N}
\end{equation}
We integrate the homogeneous equations to $\rho_m/U_0\simeq0.05$ and then use this averaged evolution. Solving the amplitude and pivot-matching conditions gives
\begin{equation}
 \begin{gathered}
 \mu_I=6.04675\times10^{15}\,\GeV,\quad
 U_0=1.12703\times10^{61}\,\GeV^4,\quad
 I_*=5.38359\times10^{17}\,\GeV,\\
 N_{\rm sr}=48.4433,\quad N_*=49.8082,\quad
 N_{*m}=51.1563,\quad N_{mR}=13.9795,\\
 \rho_{\rm end}=1.11041\times10^{61}\,\GeV^4,\qquad
 H_{\rm end}=7.90099\times10^{11}\,\GeV.
 \end{gathered}
 \label{eq:s5-background-numbers}
\end{equation}
The first two e-fold counts use $\epsilon_V=1$ in the slow-roll estimate and the first $\epsilon_H=1$ crossing in the evolved background, respectively. Halving the integration step changes the total expansion by $1.8\times10^{-6}$ e-folds and $n_s$ by $6.0\times10^{-8}$ within the adopted homogeneous approximation.

\section{Scalar minima, radial masses and matching}
\label{app:s5-scalar-minima}
\label{sec:inflation-vacuum}
The scalar realization allows the radial vacuum to be computed explicitly during inflation. Write $x=|\phi|^2$, $u=|S|^2$ and define $c_\phi=\kappa_\phi I^2+\zeta_\phi R$, $c_S=\kappa_SI^2+\zeta_SR$, with $R\simeq12H^2$ during slow roll. The magnitude-dependent potential is
\begin{equation}
 V_{\rm mag}=(-\mu_\phi^2+c_\phi)x+(-\mu_S^2+c_S)u
 +\lambda_p x^2+\lambda_Su^2+\lambda_{\phi S}xu,
 \qquad \lambda_p\equiv\lambda_\phi^{(\mathrm p)}.
 \label{eq:s5-mag-potential}
\end{equation}
The conditions $\partial_xV=\partial_uV=0$, evaluated at $x=v^2/2$ and $u=w_I^2/2$, give
\begin{equation}
 \mathsf A\begin{pmatrix}v^2\\w_I^2\end{pmatrix}
 =\begin{pmatrix}\mu_\phi^2-c_\phi\\\mu_S^2-c_S\end{pmatrix},\qquad
 \mathsf A=\begin{pmatrix}\lambda_p&\lambda_{\phi S}/2\\
 \lambda_{\phi S}/2&\lambda_S\end{pmatrix}.
 \label{eq:s5-vacuum-system}
\end{equation}
Here $v=v_\phi(I,R)$ and $w_I=w(I,R)$. The parameters $\mu_\phi^2$ and $\mu_S^2$ are fixed by the desired vacuum at $I=R=0$. Subtracting those vacuum equations eliminates both quadratic coefficients:
\begin{equation}
 \begin{pmatrix}\delta v^2\\\delta w^2\end{pmatrix}
 =-\mathsf A^{-1}\begin{pmatrix}c_\phi\\c_S\end{pmatrix},\qquad
 \mathsf A^{-1}=\frac1{D}
 \begin{pmatrix}\lambda_S&-\lambda_{\phi S}/2\\
 -\lambda_{\phi S}/2&\lambda_p\end{pmatrix},\quad
 D=\lambda_p\lambda_S-\frac{\lambda_{\phi S}^2}{4}.
 \label{eq:s5-vacuum-shifts}
\end{equation}
Positive $D$, positive diagonal quartics and positive displaced squared values give a stable interior radial minimum. The radial Hessian follows by differentiating with respect to the canonically normalized fluctuations $h$ and $s$:
\begin{equation}
 \mathsf M_{\rm rad}^2=
 \begin{pmatrix}2\lambda_pv^2&\lambda_{\phi S}vw_I\\
 \lambda_{\phi S}vw_I&2\lambda_Sw_I^2\end{pmatrix}.
 \label{eq:s5-hessian}
\end{equation}
For entries $A_r,B_r,D_r$, its eigenvalues are
\begin{equation}
 m_\pm^2=\frac{A_r+D_r\pm\sqrt{(A_r-D_r)^2+4B_r^2}}{2}.
 \label{eq:s5-eigenvalues}
\end{equation}
These masses control whether the radial fields can follow the slowly moving minimum: the conditions are $m_\pm\gg H$ and sufficiently slow changes of the minimum compared with their response times.

The matching of the quartic can be seen by completing the square in $u$:
\begin{equation}
 \begin{aligned}
 V_{\rm mag}={}&(-\mu_\phi^2+c_\phi)x+\lambda_px^2
 -\frac{(\mu_S^2-c_S-\lambda_{\phi S}x)^2}{4\lambda_S}\\
 &+\lambda_S\left[u-\frac{\mu_S^2-c_S-\lambda_{\phi S}x}{2\lambda_S}\right]^2.
 \end{aligned}
 \label{eq:s5-complete-square}
\end{equation}
At the heavy radial minimum, the last term vanishes. The coefficient of $x^2$ is therefore
\begin{equation}
 \lambda_\phi=\lambda_p-\frac{\lambda_{\phi S}^2}{4\lambda_S}.
 \label{eq:s5-quartic-threshold}
\end{equation}
This finite threshold remains after the heavy radial mode has been integrated out. The same substitution gives $\kappa_{\phi,\rm eff}=\kappa_\phi-\lambda_{\phi S}\kappa_S/(2\lambda_S)$ and the corresponding expression for $\zeta_{\phi,\rm eff}$.

For $\lambda_p=0.0100005$, $\lambda_S=0.5$ and $\lambda_{\phi S}=10^{-3}$, we obtain
\begin{equation}
 D=0.005,\qquad
 \mathsf A^{-1}=\begin{pmatrix}100&-0.1\\-0.1&2.0001\end{pmatrix},\qquad
 \lambda_\phi=0.01.
 \label{eq:s5-matrix-numbers}
\end{equation}
With $\kappa_\phi=\kappa_S=10^{-18}$ and $\zeta_\phi=\zeta_S=10^{-3}$, the pivot source is
\begin{equation}
 c_\phi=c_S=10^{-18}I_*^2+12\times10^{-3}H_*^2
 =7.60255\times10^{21}\,\GeV^2.
 \label{eq:s5-source-number}
\end{equation}
Multiplying by $-\mathsf A^{-1}$ gives
\begin{equation}
 \begin{gathered}
 \delta v^2=-7.59495\times10^{23}\,\GeV^2,\qquad
 \delta w^2=-1.44456\times10^{22}\,\GeV^2,\\
 \frac{\delta v^2}{v_\phi^2}=-7.59495\times10^{-11},\qquad
 \frac{\delta w^2}{w^2}=-2.08017\times10^{-14},\\
 m_-=1.41421\times10^{16}\,\GeV,\qquad
 m_+=8.33333\times10^{17}\,\GeV.
 \end{gathered}
 \label{eq:s5-vacuum-numbers}
\end{equation}
Both squared vacuum values stay positive. The ratios $m_-/H_*\simeq1.78\times10^4$ and $m_+/H_*\simeq1.05\times10^6$ support the adiabatic elimination of the radial modes in this example.

The change of the inflaton potential can also be obtained exactly at tree level for fixed $R$. Let $\boldsymbol y_0=(v_\phi^2,w^2)^T$ and $\boldsymbol c=(c_\phi,c_S)^T$. Evaluating the quadratic form at its displaced minimum gives
\begin{equation}
 V_{\min}(\boldsymbol c)-V_{\min}(0)
 =\frac12\boldsymbol y_0^T\boldsymbol c
 -\frac14\boldsymbol c^T\mathsf A^{-1}\boldsymbol c.
 \label{eq:s5-minimized-energy}
\end{equation}
For the inflaton dependence at fixed curvature, its leading term is
\begin{equation}
 \Delta U_{\rm tree}=\frac12(\kappa_\phi v_\phi^2+\kappa_Sw^2)I^2+\cdots.
 \label{eq:s5-tree-correction}
\end{equation}
Its slope and curvature at the pivot give
\begin{equation}
 \frac{|\Delta U'_{\rm tree}|}{|U'|}=7.18\times10^{-5},\qquad
 \frac{\mpl^2\Delta U''_{\rm tree}}{U}\simeq3.71\times10^{-7}.
 \label{eq:s5-tree-numbers}
\end{equation}
In the second expression we use $U\simeq3\mpl^2H_*^2$. The curvature couplings also shift the coefficient of the Einstein term. The measured $\mpl$ includes its constant vacuum contribution; the additional change is $\delta M_{\rm eff}^2=-\zeta_\phi\delta v^2-\zeta_S\delta w^2$, which is negligible for the values above.

\section{Heavy-particle loop corrections}
\label{app:s5-loops}
The one-loop effective potential for slowly varying masses has the form \cite{EffectivePotential}
\begin{equation}
 \Delta U=\sum_j\frac{n_jm_j^4(I)}{64\pi^2}
 \left[\ln\frac{m_j^2(I)}{Q^2}-c_j\right].
 \label{eq:s5-CW}
\end{equation}
The signed multiplicity is positive for bosons and negative for fermions. The subtraction constants are $c_j=3/2$ for scalars and fermions and $5/6$ for a massive vector. Write $m_j^2(I)=m_{j0}^2+a_jI^2+\cdots$. Since
\begin{equation}
 \frac{\dd}{\dd x}\left[x^2\left(\ln\frac{x}{Q^2}-c_j\right)\right]
 =2x\left[\ln\frac{x}{Q^2}-c_j+\frac12\right],
 \label{eq:s5-CW-derivative}
\end{equation}
the coefficient of $I^2/2$ is
\begin{equation}
 \Delta m_I^2=\sum_j\frac{n_ja_jm_{j0}^2}{16\pi^2}
 \left[\ln\frac{m_{j0}^2}{Q^2}-c_j+\frac12\right].
 \label{eq:s5-loop-mass}
\end{equation}
To compute the coefficients, differentiate the displaced vacuum with respect to $I^2$ at fixed $R$:
\begin{equation}
 \begin{pmatrix}d_v\\d_w\end{pmatrix}
 \equiv\frac{\partial}{\partial I^2}
 \begin{pmatrix}v^2\\w_I^2\end{pmatrix}
 =-\mathsf A^{-1}\begin{pmatrix}\kappa_\phi\\\kappa_S\end{pmatrix}
 =\begin{pmatrix}-9.99\times10^{-17}\\-1.9001\times10^{-18}\end{pmatrix}.
 \label{eq:s5-mass-responses}
\end{equation}
For the vector and the two Yukawa masses, $M_X^2=e_X^2(v^2+w_I^2)$ and $m_F^2=y_F^2v^2/2$, $m_\Psi^2=y_\Psi^2v^2/2$. Hence
\begin{equation}
 a_X=e_X^2(d_v+d_w)=-3.66480\times10^{-17},\qquad
 a_F=a_\Psi=\frac{0.2^2}{2}d_v=-1.9980\times10^{-18}.
 \label{eq:s5-vector-fermion-responses}
\end{equation}
Differentiating Eq.~\eqref{eq:s5-eigenvalues} gives the radial coefficients. If $\Delta_r=\sqrt{(A_r-D_r)^2+4B_r^2}$, then
\begin{equation}
 a_\pm=\frac12\left[A_r'+D_r'
 \pm\frac{(A_r-D_r)(A_r'-D_r')+4B_rB_r'}{\Delta_r}\right],
 \label{eq:s5-radial-responses}
\end{equation}
where primes here mean derivatives with respect to $I^2$. The matrix entries give $A_r'=2\lambda_pd_v$, $D_r'=2\lambda_Sd_w$ and $B_r'=B_r(d_v/v^2+d_w/w_I^2)/2$. Numerically $a_-\simeq-1.9980\times10^{-18}$ and $a_+\simeq-1.9002\times10^{-18}$.

Taking the logarithmic brackets in Eq.~\eqref{eq:s5-loop-mass} to have magnitudes of order one, the absolute numerator contributions from the light radial mode, heavy radial mode, vector, doublet and singlet fermion are approximately
\begin{equation}
 \begin{aligned}
 \sum_j|n_ja_jm_j^2|\simeq\bigl(&4.00\times10^{14}+1.32\times10^{18}
 +2.79\times10^{19}\\
 &+3.20\times10^{15}+1.60\times10^{15}\bigr)\,\GeV^2.
 \end{aligned}
 \label{eq:s5-loop-contributions}
\end{equation}
The multiplicities are $1,1,3,-8,-4$. Dividing the sum by $16\pi^2$ gives $1.85\times10^{17}\,\GeV^2$, corresponding to $|\Delta\eta_V|\sim9.73\times10^{-8}$. This is an order-of-magnitude estimate with matching scales chosen near the relevant masses. Counterterms and logarithms must be specified for a renormalized prediction. The larger correction from the visible decay interaction and kinetic mixing is calculated separately below.

\section{Phase normalization and isocurvature}
\label{app:s5-isocurvature}
At a slowly changing radial minimum, $F_\chi=vw_I/\sqrt{v^2+w_I^2}$. Its logarithmic variation is
\begin{equation}
 \delta\ln F_\chi=\frac12\left[
 \frac{w^2}{v_\phi^2+w^2}\frac{\delta v^2}{v_\phi^2}
 +\frac{v_\phi^2}{v_\phi^2+w^2}\frac{\delta w^2}{w^2}\right]
 =-3.74\times10^{-11}.
 \label{eq:s5-F-shift}
\end{equation}
After integrating out the vector, the angular action is $\tfrac12F_\chi^2(t)(\partial\theta)^2-V(\theta)$, where $\theta=-\chi/F_\chi$ for constant vacuum values. Its homogeneous equation is
\begin{equation}
 \frac{\dd}{\dd t}\left(a^3F_\chi^2\dot\theta\right)
 +a^3\frac{\partial V}{\partial\theta}=0.
 \label{eq:s5-angle-eom}
\end{equation}
When the angular force is negligible during inflation, $a^3F_\chi^2\dot\theta$ is conserved. Expansion therefore damps an initial angular velocity and leaves an almost fixed angle.

A canonically normalized light scalar has fluctuation amplitude $H_*/(2\pi)$ per logarithmic interval of wave number. For an adiabatically varying $F_\chi$, this gives $\delta\theta=H_*/(2\pi F_{\chi,*})$. The late density responds through
\begin{equation}
 \frac{\delta\rho_{\rm coh}}{\rho_{\rm coh}}
 =D_\theta\delta\theta,\qquad
 D_\theta\equiv\frac{\partial\ln\rho_{\rm coh}}{\partial\theta_i}.
 \label{eq:s5-density-response}
\end{equation}
For a fraction $f_{\rm coh}$ of the dark matter, uncorrelated with the inflaton fluctuations, the isocurvature power is
\begin{equation}
 \mathcal P_S=f_{\rm coh}^2D_\theta^2
 \left(\frac{H_*}{2\pi F_{\chi,*}}\right)^2,\qquad
 \beta_{\rm iso}=\frac{\mathcal P_S}{\mathcal P_{\mathcal R}+\mathcal P_S}.
 \label{eq:s5-iso-power}
\end{equation}
The response $D_\theta$ can be reproduced from the cosine evolution during radiation domination. With $x=m_{\chi,0}t$ and $H=1/(2t)$, the equation and the late density invariant are
\begin{equation}
 \theta''+\frac{3}{2x}\theta'+\sin\theta=0,\qquad
 \mathcal J(\theta_i)=\left\langle
 x^{3/2}\left(\frac12\theta'^2+1-\cos\theta\right)\right\rangle.
 \label{eq:s5-cosine-response}
\end{equation}
Primes in this equation mean $\dd/\dd x$. The regular initial solution is $\theta=\theta_i-\sin\theta_i\,x^2/5+\cdots$. Since the final density is proportional to $\mathcal J$, finite differences give
\begin{equation}
 D_\theta\simeq
 \frac{\ln\mathcal J(\theta_i+\delta)-\ln\mathcal J(\theta_i-\delta)}{2\delta}
 =6.20353,
 \quad\theta_i=0.3252048,\quad\delta=10^{-4}.
 \label{eq:s5-response-number}
\end{equation}
We integrate to $x=1000$ and average over $800<x<1000$. The harmonic estimate $2/\theta_i\simeq6.15$ explains the size of this derivative.

For $f_{\rm coh}=1$, the pivot values give $\mathcal P_S=6.26454\times10^{-11}$ and $\beta_{\rm iso}=0.028967$. Solving $\beta_{\rm iso}<\beta_{\max}$ for the inflationary scale gives
\begin{equation}
 H_*<\frac{2\pi F_{\chi,*}}{f_{\rm coh}|D_\theta|}
 \sqrt{\frac{\beta_{\max}}{1-\beta_{\max}}\mathcal P_{\mathcal R}}
 =9.1590\times10^{11}\,\GeV
 \quad(\beta_{\max}=0.038).
 \label{eq:s5-Hbound}
\end{equation}
This uses the uncorrelated isocurvature limit of Ref.~\cite{Inflation}. The power also lies below the quoted $1.1\times10^{-10}$ amplitude bound in the combined analysis of Ref.~\cite{ACTInflation}, under its scale-invariance assumptions. Over sixty e-folds with nearly constant $H/F_\chi$, independent long-wavelength increments give $\langle\delta\theta^2\rangle^{1/2}\simeq\sqrt{60}H_*/(2\pi F_\chi)=9.88\times10^{-6}$.

Repeating the background matching for nearby seed scales gives $H_*\simeq5.9695\times10^{11}\,\GeV$, $n_s\simeq0.96932$, $\beta_{\rm iso}\simeq0.01650$, and $H_*\simeq6.9644\times10^{11}\,\GeV$, $n_s\simeq0.96936$, $\beta_{\rm iso}\simeq0.02233$. These values reproduce the additional examples quoted in the paper. Comparing $n_s=0.96940$ with the marginalized values $0.9709\pm0.0038$ and approximately $0.974\pm0.003$ quoted in Ref.~\cite{ACTBase} gives offsets of about $0.4$ and $1.5$ standard deviations, respectively. Such comparisons retain the published cosmological assumptions; a model likelihood would fit the scalar and isocurvature spectra jointly.

\section{Perturbative reheating and inflaton fluctuations}
\label{app:s5-reheating}
\label{sec:inflation-reheating}
Expanding the potential near $I=0$ gives
\begin{equation}
 U=\frac{U_0}{\mu_I^2}I^2-\frac{U_0}{\mu_I^4}I^4+\cdots
 =\frac12m_I^2I^2+\frac{\lambda_4}{24}I^4+\cdots,
 \quad m_I^2=\frac{2U_0}{\mu_I^2},\quad
 \lambda_4=-\frac{24U_0}{\mu_I^4}.
 \label{eq:s5-origin-expansion}
\end{equation}
The benchmark gives $m_I=7.85164\times10^{14}\,\GeV$ and $\lambda_4=-0.20233$. The exact potential is bounded below even though this local quartic term is negative.

Let the visible decay interaction be $-IF_Y^2/(4\mathcal M_Y)$. For two massless gauge bosons, the vertex is proportional to $(k_1\cdot k_2)g^{\mu\nu}-k_1^\nu k_2^\mu$. Summing transverse polarizations gives $\sum|\mathcal A|^2=m_I^4/(2\mathcal M_Y^2)$. The two-body phase space, including the identical-particle factor, gives
\begin{equation}
 \Gamma_{I\to YY}=\frac{\sum|\mathcal A|^2}{32\pi m_I}
 =\frac{m_I^3}{64\pi\mathcal M_Y^2}.
 \label{eq:s5-decay-width}
\end{equation}
At the end of reheating, the approximation $H_R=\Gamma_I$ and $\rho_R=\pi^2g_{*,R}T_R^4/30$ gives
\begin{equation}
 \Gamma_I=\sqrt{\frac{\pi^2g_{*,R}}{90}}\frac{T_R^2}{\mpl},\qquad
 T_R=\left(\frac{90}{\pi^2g_{*,R}}\right)^{1/4}\sqrt{\Gamma_I\mpl}.
 \label{eq:s5-TR}
\end{equation}
For visible-dominated reheating at $10^{10}\,\GeV$, the results are $\Gamma_I=140.512\,\GeV$, $\mathcal M_Y=1.30894\times10^{20}\,\GeV$ and $\mpl/\mathcal M_Y=0.018603$. The pivot expansion parameter is $I_*/\mathcal M_Y=0.004113$.

The oscillating gauge kinetic coefficient changes by $q_Y=\mathcal I_{\rm amp}/\mathcal M_Y$. The total end energy bounds an amplitude through $U(\mathcal I_{\rm amp})\leq\rho_{\rm end}$, giving
\begin{equation}
 \mathcal I_{\rm amp}\leq\mu_I
 \sqrt{\frac{\rho_{\rm end}/U_0}{1-\rho_{\rm end}/U_0}}
 =4.9423\times10^{16}\,\GeV.
 \label{eq:s5-amplitude-bound}
\end{equation}
Thus $q_Y\lesssim3.7758\times10^{-4}$ and $q_Y^2m_I/H_{\rm end}\lesssim1.4168\times10^{-4}$. In a narrow resonance band, the growth rate is of order $q_Ym_I$ and expansion sweeps a mode across the fractional band width in a time of order $q_Y/H$. Their product gives the estimate $q_Y^2m_I/H$ \cite{ReheatingResonance}. The estimate supports the perturbative treatment of this visible decay channel. The displayed heavy hidden particles are kinematically inaccessible in two-body inflaton decay: $m_I/(2m_F)=0.02776$.

The inflaton itself can undergo self-resonance. Linearizing its equation around the homogeneous solution gives, for modes well inside the Hubble radius,
\begin{equation}
 \delta\ddot I_k+3H\delta\dot I_k
 +\left(\frac{k^2}{a^2}+U''(I)\right)\delta I_k=0.
 \label{eq:s5-self-resonance}
\end{equation}
Equation~\eqref{eq:s5-potential-derivatives} gives $U''<0$ for $|I|>\mu_I/\sqrt3$. Modes with sufficiently small physical momentum can therefore grow during this part of an oscillation. A nonlinear calculation is needed to determine fragmentation and the resulting pressure \cite{InflatonFragmentation}. The homogeneous calculation estimates this uncertainty by replacing the late averaged pressure with $\overline w_I=1/3$ and repeating the amplitude and expansion matching. It then gives $N_*=53.3147$, $\mu_I=6.71157\times10^{15}\,\GeV$ and $n_s=0.971433$. This replacement is a sensitivity calculation for a prescribed pressure history.

\section{Kinetic mixing and the inflaton slope}
\label{app:s5-kinetic-mixing}
The visible decay operator also affects the heavy vector through hypercharge--portal kinetic mixing. Define $A_Y(I)=1+I/\mathcal M_Y$. The quadratic kinetic matrix for the hypercharge and portal vectors is
\begin{equation}
 \mathsf K=\begin{pmatrix}A_Y&\epsilon\\\epsilon&1\end{pmatrix},
 \qquad \mathsf M^2=\begin{pmatrix}0&0\\0&M_X^2\end{pmatrix}.
 \label{eq:s5-kinetic-matrix}
\end{equation}
The physical poles solve $\det(m^2\mathsf K-\mathsf M^2)=0$. Expanding the determinant gives $m^2[m^2(A_Y-\epsilon^2)-A_YM_X^2]=0$, so the massive pole is
\begin{equation}
 m_{X,\rm pole}^2(I)=\frac{A_YM_X^2}{A_Y-\epsilon^2},\qquad
 \frac{\dd m_{X,\rm pole}^2}{\dd I}
 =-\frac{M_X^2\epsilon^2}{\mathcal M_Y(A_Y-\epsilon^2)^2}.
 \label{eq:s5-vector-pole}
\end{equation}
The kinetic matrix is positive for $A_Y>\epsilon^2$. Inserting this mass into the vector part of Eq.~\eqref{eq:s5-CW}, with multiplicity three, gives
\begin{equation}
 \Delta U_Y'=\frac{3}{32\pi^2}m_{X,\rm pole}^2
 \frac{\dd m_{X,\rm pole}^2}{\dd I}
 \left[\ln\frac{m_{X,\rm pole}^2}{Q^2}-\frac13\right].
 \label{eq:s5-vector-loop-slope}
\end{equation}
For $A_Y\simeq1$, small $\epsilon$ and $Q\simeq M_X$, this gives
\begin{equation}
 |\Delta U_Y'|\simeq\frac{\epsilon^2M_X^4}{32\pi^2\mathcal M_Y},\qquad
 \frac{|\Delta U_Y'|}{|U'|}\simeq294.6
 \left(\frac{\epsilon}{10^{-3}}\right)^2.
 \label{eq:s5-mixing-slope-number}
\end{equation}
Keeping this uncancelled estimate below one percent of the required slope gives $|\epsilon|\lesssim5.83\times10^{-6}$. The actual renormalized slope includes the local counterterms. The adopted plateau therefore requires adjustment of those terms or a mechanism protecting its shape.

Visible fermions generate kinetic mixing because they carry both hypercharge and $B-L$. The charge product is unchanged by charge conjugation, so one can sum the physical left- and right-handed charges. In one generation,
\begin{equation}
 \begin{aligned}
 \sum Y(B-L)={}&6\left(\frac16\right)\left(\frac13\right)
 +3\left(\frac23\right)\left(\frac13\right)
 +3\left(-\frac13\right)\left(\frac13\right)\\
 &+2\left(-\frac12\right)(-1)+(-1)(-1)+0
 =\frac83.
 \end{aligned}
 \label{eq:s5-charge-sum}
\end{equation}
The terms are the quark doublet, up singlet, down singlet, lepton doublet, charged lepton and right-handed neutrino. Three generations give $8$; the visible Higgs has zero $B-L$. The one-loop magnitude is consequently
\begin{equation}
 \left|\frac{\dd\epsilon}{\dd\ln\mu}\right|
 =\frac{g_Yg_X}{16\pi^2}\frac23\,8
 =\frac{g_Yg_X}{16\pi^2}\frac{16}{3}
 \simeq0.00405\quad(g_Y=0.4,\ g_X=0.3).
 \label{eq:s5-mixing-beta}
\end{equation}
This explains why a small renormalized mixing at one scale needs further control if it is to remain small over a substantial range of scales.

\section{Temperature bounds and the hidden reheating fraction}
\label{app:s5-temperature}
The radiation temperature during slow inflaton decay can exceed $T_R$ \cite{MaximumTemperature}. Its energy density is bounded above by the total energy available at the end of inflation. Equating that upper bound to a visible radiation density gives
\begin{equation}
 T_{\max}\leq\left(\frac{30\rho_{\rm end}}{\pi^2g_*}\right)^{1/4}
 =7.4987\times10^{14}\,\GeV\quad(g_*=106.75).
 \label{eq:s5-Tmax}
\end{equation}
For the conservative positive thermal sources $0\leq c_{\phi,\rm th},c_{S,\rm th}\leq T_{\max}^2$, Eq.~\eqref{eq:s5-vacuum-shifts} gives
\begin{equation}
 \frac{|\delta v^2|}{v_\phi^2}\lesssim
 \frac{100T_{\max}^2}{v_\phi^2}\simeq0.00563,\qquad
 \frac{|\delta w^2|}{w^2}\lesssim
 \frac{2.0001T_{\max}^2}{w^2}\simeq1.62\times10^{-6}.
 \label{eq:s5-thermal-shifts}
\end{equation}
Allowing the two positive sources to vary independently gives the componentwise bounds written here. Taking both equal to $T_{\max}^2$, as in the manuscript's numerical check, gives $0.005617$ and $1.539\times10^{-6}$. Both choices preserve the nonzero vacuum values. These are bounds on assumed thermal mass coefficients; a colder hidden plasma gives smaller corrections. For the second model with $\lambda_{12}=0$, $v_1=v_2=1.40414\times10^{17}\,\GeV$ and $\lambda_1=\lambda_2=0.01$, the same calculation gives $\delta v_i^2/v_i^2=-3.86\times10^{-11}$, $\delta F_2/F_2=-1.93\times10^{-11}$ and a thermal bound of $0.00285$.

Let $\xi_R=T_{D,R}/T_R$ be the hidden-to-visible temperature ratio just after reheating, assuming the hidden gluons have internally thermalized. The hidden and visible energy densities are $\rho_D=(\pi^2/30)6T_{D,R}^4$ and $\rho_{\rm vis}=(\pi^2/30)g_{*,R}T_R^4$. Therefore
\begin{equation}
 \frac{B_D}{B_{\rm vis}}=\frac{\rho_D}{\rho_{\rm vis}}
 =\frac{6}{g_{*,R}}\xi_R^4.
 \label{eq:s5-branching}
\end{equation}
The factor six counts two transverse polarizations for each of the three hidden gauge bosons. The manuscript uses $\xi_*=10^{-3}$ as a late deconfined-radiation extrapolation relative to the visible entropy temperature. Visible entropy conservation relates its normalization to reheating through
\begin{equation}
 \xi_R=\xi_*\left(\frac{g_{*s,R}}{g_{*s,0}}\right)^{1/3}
 =0.00301137,
 \qquad \frac{B_D}{B_{\rm vis}}=4.6221\times10^{-12}.
 \label{eq:s5-branching-number}
\end{equation}
This extrapolated normalization is used to specify the initial hidden entropy. The later confined population requires its own evolution equations.

For the hidden decay operator $-IW_D^aW_D^a/(4\mathcal M_D)$, the three gauge species give
\begin{equation}
 \frac{\Gamma_{I\to WW}}{\Gamma_{I\to YY}}
 =3\left(\frac{\mathcal M_Y}{\mathcal M_D}\right)^2.
 \label{eq:s5-hidden-width-ratio}
\end{equation}
If these operators dominate reheating, the required energy fraction implies
\begin{equation}
 \begin{gathered}
 \frac{\mathcal M_D}{\mathcal M_Y}\gtrsim
 \sqrt{\frac3{4.6221\times10^{-12}}}=8.0564\times10^5,
 \qquad \mathcal M_D\gtrsim1.0545\times10^{26}\,\GeV,\\
 d_Y\equiv\frac{\mpl}{\mathcal M_Y}=0.018603,\qquad
 |d_D|\equiv\frac{\mpl}{\mathcal M_D}\lesssim2.3091\times10^{-8}.
 \end{gathered}
 \label{eq:s5-decay-hierarchy}
\end{equation}
The scales $\mathcal M_Y$ and $\mathcal M_D$ parametrize inverse operator coefficients. A large value can encode a small dimensionless coupling at a lower physical cutoff. The gauge symmetries displayed in the model allow both operators, so the hierarchy of their coefficients requires additional microscopic input.

Nonlinear fragmentation and a microscopic origin of the reheating hierarchy remain separate dynamical calculations. The scalar example establishes the stated conditional evolution once its potential and couplings are specified.